\documentclass[prc,english,aps,twocolumn,nofootinbib,preprintnumbers,
showpacs,showkeys,superscriptaddress,fleqn,floatfix]{revtex4}
\usepackage[T1]{fontenc}
\usepackage[latin9]{inputenc}
\usepackage{bm}
\usepackage{amsbsy}
\usepackage{graphicx}
\usepackage{esint}

\makeatletter

\usepackage[normalem]{ulem} 
\usepackage[dvipsnames]{xcolor} 
\renewcommand\sout{\bgroup \color{red} \ULdepth=-.5ex \ULset}

\providecommand{\\}{\\}
\@ifundefined{textcolor}{}
{%
 \definecolor{BLACK}{gray}{0}
 \definecolor{WHITE}{gray}{1}
 \definecolor{RED}{rgb}{1,0,0}
 \definecolor{GREEN}{rgb}{0,1,0}
 \definecolor{BLUE}{rgb}{0,0,1}
 \definecolor{CYAN}{cmyk}{1,0,0,0}
 \definecolor{MAGENTA}{cmyk}{0,1,0,0}
 \definecolor{YELLOW}{cmyk}{0,0,1,0}
}

\makeatother

\usepackage{babel}
\begin{document}

\preprint{INHA-NTG-04/2015}
\title{Modification of generalized vector form factors and transverse 
  charge densities \\ of the nucleon in nuclear matter}

\author{Ju-Hyun Jung}
\email{juhyun@inha.edu}
\affiliation{
Department of Physics, Inha University, Incheon 22212, Republic of
Korea} 
\affiliation{Institute of Physics, University of Graz,
  Universitätsplatz 5, A-8010 Graz, Austria} 

\author{Ulugbek Yakhshiev}
\email{yakhshiev@inha.ac.kr}
\affiliation{
Department of Physics, Inha University, Incheon 22212, Republic of
Korea} 

\author{Hyun-Chul Kim}
\email{hchkim@inha.ac.kr}
\affiliation{
Department of Physics, Inha University, Incheon 22212, Republic of
Korea} 
\affiliation{School of Physics, Korea Institute for Advanced Study
  (KIAS), Seoul 02455, Republic of Korea}

\begin{abstract}
We investigate the medium modification of the generalized vector form
factors of the nucleon, which include the electromagnetic and
energy-momentum tensor form factors, based on an in-medium modified  
$\pi$-$\rho$-$\omega$ soliton model. We find that the vector form
factors of the nucleon in nuclear matter fall off faster than those in
free space, which implies that the charge radii of the nucleon become
larger in nuclear medium than in free space. We also compute the
corresponding transverse charge densities of the nucleon in nuclear
matter, which clearly reveal the increasing of the nucleon size in
nuclear medium.  
\end{abstract}




\maketitle


\section{Introduction}
Understanding the electromagnetic form factors (EMFFs) of the nucleon
has been one of the most important issues in hadronic physics, since
they reveal the internal quark structure of the nucleon. While the
EMFFs of the nucleon have been studied well over several decades,
their precise data were obtained only recently by measuring the
transverse and longitudinal recoil proton
polarisations~\cite{Jones:1999rz, Gayou:2001qt,Gayou:2001qd, 
Punjabi:2005wq, Puckett:2010ac, Bernauer:2010wm, Ron:2011rd,
  Zhan:2011ji, Puckett:2011xg, Schlimme:2013eoz,
  Bernauer:2013tpr}. These new experimental data have drawn a great
deal of attention both experimentally and 
theoretically (see recent reviews and references 
therein~\cite{HydeWright:2004gh,Arrington:2006zm,Perdrisat:2006hj, 
  Pacetti:2015iqa}). In the meanwhile, form factors of the nucleon can
be defined as Mellin moments of the corresponding generalized parton
distributions (GPDs) that unveil novel aspect of the internal
structure of the nucleon~\cite{Mueller:1998fv,Ji:1996ek, Ji:1996nm,
  Radyushkin:1996nd} (see also the following
reviews~\cite{Goeke:2001tz, Diehl:2003ny, Belitsky:2005qn}). 
This new definition of the form factors enables one to get access to
the energy-momentum tensor form factors (EMTFFs) and the tensor form
factors via the GPDs, which cannot be otherwise directly measured 
experimentally. In this definition, the energy-momentum tensor form
factors can be also understood as the second Mellin moments of the
isoscalar vector GPDs of the nucleon. 

The Fourier transforms of the generalized vector form factors of the
nucleon including the EMFFs and the EMTFFs in the transverse plane, as
viewed from a light front frame moving towards a nucleon, makes it
possible to see how the charge densities of quarks are 
distributed transversely~\cite{Burkardt:2002hr,Burkardt:2000za}. 
These are called transverse charge densities and they provide correctly a
probability of finding quarks inside a nucleon in the transverse
plane.  Transverse charge densities have been already investigated for the
unpolarized~\cite{Miller:2007uy} and transversely
polarized~\cite{Carlson:2007xd} nucleons. 

Furthermore, it is of equal importance to examine
how the EM structure of the nucleon is changed in nuclear
matter. Studying the EMFFs of the nucleon in nuclear medium
provides a new perspective on EM properties of the nucleon modified in  
nuclei~\cite{Malace:2008gf,Dieterich:2000mu,Avakian:2003pk,
Strauch:2002wu,Lu:1998tn,Smith:2004dn,Cloet:2009tx,
Yakhshiev:2002sr,Yakhshiev:2012zz}. 
In fact, the first experimental study of deeply virtual
Compton scattering on (gaseous) nuclear targets (H, He, N, Ne, Kr, Xe)
was reported in Ref.~\cite{Airapetian:2009bi}. While uncertainties of 
the first measurement are so large that one is not able to observe
nuclear modifications of the nucleon structure, future experiments
will provide more information on medium modifications of the EM
properties of the nucleons.

In the present work, we want to investigate the nucleon EMFFs and the
transverse charge densities of quarks inside a nucleon in nuclear
matter within the framework of an in-medium 
modified soliton model with explicit $\pi$-$\rho$-$\omega$
degrees of freedom. The model has certain virtues: it is
simple but respects the chiral symmetry and its spontaneous
breaking. Moreover, one can easily extend it including the influence of 
surrounding nuclear environment to the nucleon properties
 based on modifications of the meson properties in nuclear 
medium~\cite{Rakhimov:1996vq,Yakhshiev:2010kf}. In this context, 
the EMTFFs of the nucleon, which are yet another fundamental 
form factors that are related to the generalized EMFFs, have
been investigated in free space~\cite{Cebulla:2007ei,Jung:2013bya} 
and in nuclear matter within the chiral soliton 
approaches~\cite{Kim:2012ts,Jung:2014jja}. 
The results have explained certain interesting features of the
modifications of nucleon properties in nuclear 
matter such as the pressure and angular momentum.
Indeed, we will also show in this work how
the EM properties of the nucleon are changed in nuclear matter in a
simple manner. We will also see that the transverse charge densities
expose noticeably how the distribution of quarks undergo
changes in the presence of nuclear medium.

The present paper is organized as follows: In Sec. II, we briefly
explain the general formalism of the $\pi$-$\rho$-$\omega$ soliton
model modified in nuclear medium. In Sec. III, we describe how one can
compute the generalized vector form factors within this framework. In
Sec. IV, we present the results from the present work and discuss
them. The final Section is devoted to the summary and the conclusion.   

\section{General formalism}
We start from the in-medium modified effective chiral Lagrangian with
the $\pi$, $\rho$, and $\omega$ meson degrees of freedom, where
the nucleon arises as a topological soliton~\cite{Jung:2012sy}. 
The Lagrangian has the form
\begin{eqnarray}
\mathcal{L}^{*} & = & \mathcal{L}_{\pi}^{*}+\mathcal{L}_{V}^{*}
+\mathcal{L}_{\mathrm{kin}}^{*}+\mathcal{L}_{\mathrm{WZ}}^{*},
\label{Lag}
\end{eqnarray}
where the corresponding terms are expressed as 
\begin{eqnarray}
\mathcal{L}_{\pi}^{*} & = & \frac{f_{\pi}^{2}}{4}\,
\mbox{Tr}\left(\partial_{0}U\partial_{0}U^{\dagger}\right)
-\alpha_{p}\frac{f_{\pi}^{2}}{4}\,\mbox{Tr}
\left(\partial_{i}U\partial_{i}U^{\dagger}\right)
\cr&&
+\,\alpha_{s}\frac{f_{\pi}^{2}m_{\pi}^{2}}{2}\,\mbox{Tr}\left(U-1\right)\,,
\label{begLag}\\
\mathcal{L}_{V}^{*} & = & \frac{f_{\pi}^{2}}{2}\,
\mbox{Tr}\left[D_{\mu}\xi\cdot\xi^{\dagger}
+D_{\mu}\xi^{\dagger}\cdot\xi\right]^{2}\,,
\label{LV}\\
\mathcal{L}_{\mathrm{kin}}^{*} & = & 
-\frac{1}{2g_{V}^{2}\zeta_V}\,\mbox{Tr}\left(F_{\mu\nu}^{2}\right)\,,
\label{kin}\\
\mathcal{L}_{\mathrm{WZ}}^{*} & = & \left(\frac{N_{c}}{2}g_{\omega}
  \sqrt{\zeta_\omega}\right)\omega_{\mu}
\frac{\epsilon^{\mu\nu\alpha\beta}}{24\pi^{2}}  \cr
&&\times \,\mbox{Tr} \left\{ \left(U^{\dagger} \partial_{\nu} U
  \right) \left(U^{\dagger}\partial_{\alpha}U\right)
  \left(U^{\dagger}\partial_{\beta}U\right)\right\}. 
\label{endLag}
\end{eqnarray}
Here the asterisk designates medium modified quantities. 
The SU(2) chiral field is written as $U=\xi_{L}^{\dagger}\,\xi_{R}$ in
unitary gauge, and the field-strength tensor and the covariant
derivative are defined, respectively, as   
\begin{eqnarray}
F_{\mu\nu} & = & \partial_{\mu} V_{\nu} - \partial_{\nu} V_{\mu} -
i[V_{\mu},V_{\nu}]\,,
\label{Fmu}\\  
D_{\mu}\,\xi_{L(R)} & = & \partial_{\mu}\, \xi_{L(R)}-i\, V_{\mu}\,\xi_{L(R)},
\label{Dmu}
\end{eqnarray}
where the vector field $V_\mu$ includes the $\rho$-meson and 
$\omega$-meson fields, i.e. $\bm \rho_\mu $ and $\omega_\mu$,
respectively, expressed as
\begin{equation}
V_\mu \;=\; \frac{g_V\sqrt{\zeta_V}}{2} (\bm \tau \cdot \bm \rho_\mu +
\omega_\mu) 
\label{Vmu}   
\end{equation}
with the Pauli matrices $\bm \tau$ in isospin space.

Note that in Eqs.~(\ref{LV}), (\ref{kin}) and (\ref{Vmu}) 
the subscript $V$ generically stands for
both the $\rho$-meson and the $\omega$-meson and for compactness
we keep the generic form of those expressions. One can separate
Eqs.~(\ref{LV}) and $(\ref{kin})$ into the $\rho$- and $\omega$-meson
parts using the definitions~(\ref{Fmu}), (\ref{Dmu}) and (\ref{Vmu}). 
Then $g_V$ designates $g_\rho$ for the $\rho$ meson or $g_\omega$ for
the $\omega$. $N_c=3$ is the number of colors. 

The input parameters of the model in
Eqs.~(\ref{begLag})-(\ref{endLag}) can be classified into two
different classes:  the parameters $f_\pi$, $m_\pi$, $g_\rho$,
$g_\omega$ and $N_c$ are related to the corresponding observables in
free space, while $\alpha_p$, $\alpha_s$ and $\zeta_V$ are
pertinent to properties of pionic atoms and infinite and homogenous 
nuclear matter~\footnote{$\zeta_V$ 
 denotes also a generic form for both $\zeta_\rho$ and $\zeta_\omega$ 
 which appear in the corresponding $\rho$- and $\omega$-meson parts
 of the Lagrangian.}.

In free space, in-medium parameters are all set equal to one:
$\alpha_p=\alpha_s=\zeta_\omega=\zeta_\rho=1$  and the other
parameters are fixed by using either experimental or empirical data on  
the pion and the vector mesons~\cite{Meissner:1986hi}. 
The pion decay constant and mass are taken to be $f_{\pi}=93$~MeV and
$m_{\pi}=135$~MeV (the neutral pion mass).    
The values of the coupling constants for the $\rho$ and $\omega$
mesons are given respectively as $g_{\rho}=5.86$ and
$g_{\omega}=5.95$. The Kawarabayashi-Suzuki--Riazuddin-Fayyazuddin
(KSRF) relation connects them to the masses of the vector mesons,  
i.e. \ $m_{\rho}=770$~MeV and $m_{\omega}=782$~MeV, as follows  
\begin{eqnarray}
2f_{\pi}^{2}g_{\rho}^{2}=m_{\rho}^{2}\,,\qquad 2f_{\pi}^{2}g_{\omega}^{2}
=m_{\omega}^{2}\,.
\end{eqnarray}

In general, the parameters $\alpha_p$,
$\alpha_s$, and $\zeta_V$ stand for the medium functionals which are the
essential quantities in the present work.  
They depend on the nuclear matter density $\rho$ and are defined as
\begin{eqnarray}
\alpha_{p}(\rho)  &=&1-\frac{4\pi c_{0}\rho/\eta}{1+g_{0}'4\pi
  c_{0}\rho)/\eta}\,,\cr 
\alpha_{s} (\rho) &=&1-{4\pi\eta b_{0}\rho}{m_{\pi}^{-2}}\,,\cr
\zeta_V(\rho) &=& \exp\left\{ -\frac{\gamma_{{\rm num}}\rho}
  {1+\gamma_{{\rm den}}\rho}\right\}  \,.
\label{medfunc}
\end{eqnarray}
They provide crucial information on how the nuclear-matter environment
influences properties of the single soliton~\cite{Jung:2012sy}. 
The $\eta$ is a kinematic factor defined as
$\eta=1+m_{\pi}/m_{N}\simeq1.14$. The values of the empirical parameters 
$b_{0}=-0.024\, m_{\pi}^{-1}$ and $c_{0}=0.09\, m_{\pi}^{-3}$ are taken from the
analysis of pionic atoms and the data on low-energy pion-nucleus
scattering. The $g_{0}'=0.7$ denotes the Lorentz-Lorenz factor that takes
into account the short-range correlations~\cite{Ericsonbook}. 

The additional parameters  $\gamma_{\rm num}$ and $\gamma_{\rm den}$ 
are introduced phenomenologically to reproduce the saturation 
point at normal nuclear matter. Two different models have been
discussed in the framework of the present approach~\cite{Jung:2012sy},
in order to introduce a nuclear modification in the present soliton
approach, which we will briefly explain here. In {\it Model I}, one
neglects the small mass difference of the $\rho$ and $\omega$ mesons
in free space ($m_\omega=m_\rho=770$~MeV,  $g_\omega=g_\rho=5.86$) and
assumes that the KSRF relation still holds in nuclear matter  
\begin{eqnarray}
2f_{\pi}^{2}g_{\rho}^{2}\zeta_\rho&=& m_\rho^{*2} \;=\;
m_\omega^{*2},\quad \zeta_\rho=\zeta_\omega\neq 1. 
\label{eq:KSRF1}
\end{eqnarray}
In {\it Model II}, on the other hand, we remove the degeneracy of the
vector meson masses in free space ($m_\rho\neq m_\omega=782$~MeV,
$g_\rho\neq  g_\omega=5.95$), and instead of Eq.~(\ref{eq:KSRF1})
assume that the KSRF relation is valid only for the $\rho$ meson, with
the $\omega$ meson kept as in free space: 
\begin{eqnarray}
2f_{\pi}^{2}g_{\rho}^{2}\zeta_\rho&=& m_\rho^{*2} \;\neq\;
m_\omega^{*2}, \quad \zeta_\rho \neq 1, \;\;\;\; \zeta_\omega =1.
\label{eq:KSRF2}
\end{eqnarray}

These two different models are devised to implement possible ways of
nuclear modification. We take into account the possibility that the
$\rho$ and $\omega$ meson degrees of freedom could respond differently
to a nuclear environment~\cite{Naruki:2005kd,Wood:2008ee}. 
The effects of the $\omega$-mesons are mainly limited to the inner  
core of the nucleon. Therefore, the two variants of the model describe
the situation that the inner core of the nucleon is more (Model I) 
or less (Model II) affected by medium effects. The latter is
a plausible scenario, at least around the normal nuclear 
matter density.

In practice, these two models yield comparable results 
in many respects. A notable (and in our context important) 
difference, however, is the description of the incompressibility  
of symmetric nuclear matter: 
Model I yields a smaller value of the incompressibility, while 
Model II produces a larger one. It means that  
Model II gives a stiffer nuclear binding energy and agrees better  
with the data (see explanations in Ref.~\cite{Jung:2012sy}). In both
models the values of $\gamma_{\mathrm{num}}$ and
$\gamma_{\mathrm{den}}$ are fitted to reproduce the coefficient of the
volume term in the empirical mass 
formula $a_V\approx 26$~MeV. Although this is larger than the
experimental value $a_V^{\rm exp}\approx 16$~MeV, the relative change  
of the in-medium nucleon mass is reproduced correctly (See Eq.~(12) in
Ref.~\cite{Jung:2012sy} and the corresponding explanation.). In Model
I we have $\gamma_{\mathrm{num}}=2.390\,m_{\pi}^{-3}$ and
$\gamma_{\mathrm{den}}=1.172\, m_{\pi}^{-3}$, whereas in Model II we
employ $\gamma_{\mathrm{num}}=1.970\, m_{\pi}^{-3}$ and
$\gamma_{\mathrm{den}}=0.841\, m_{\pi}^{-3}$. For further details on
these two models in relation to nuclear matter properties, and 
to the classical and quantum solution in free space and in nuclear matter 
we refer to Refs.~\cite{Jung:2012sy,Jung:2014jja}. 
In the next section we concentrate on generalized form factors and the
corresponding transverse charge densities.

\section{Generalized vector form factors and 
transverse charge densities}  
The generalized vector form factors of the nucleon can be defined as
the matrix element of a vector operator as follows:
\begin{widetext}
  \begin{eqnarray}
\langle N(p',s)| \bar{\psi}(0) \gamma^{\{} i D^{\mu_1}\cdots i
 D^{\mu_n\}}\psi(0) |N(p,s)\rangle 
&=& \bar{u}(p',s')\left[
    \sum_{i=0,\mathrm{even}}^n\left\{\gamma^{\{\mu}
    \Delta^{\mu_1}\cdots \Delta^{\mu_i} P^{\mu_i+1}\cdots P^{\mu_n\}}
    A_{n+1,i}(\Delta^2) \right. \right. \cr 
&& \left.\left.\hspace{-7cm}- i \frac{\Delta_\alpha
    \sigma^{\alpha\{\mu}}{2M_N} \Delta^{\mu_1}\cdots \Delta^{\mu_i}
    \bar{P}^{\mu_i+1}\cdots \bar{P}^{\mu_n\}} B_{n+1,i}(\Delta^2)
   \right\} + \left. \frac{\Delta^{\mu} \Delta^{\mu_1}\cdots \Delta^{\mu_n}}{M_N}
   C_{n+1,0}(\Delta^2)\right|_{n\,\mathrm{odd}}
\right] u(p,s),
 \label{eq:GFF}
  \end{eqnarray}
\end{widetext}
where $D^\mu_i$ denotes the covariant operator in quantum
chromodynamics (QCD) and the braces $``\{\}"$ stand for the
symmetrization. Here $\Delta^{\mu_i}$ and the $P^{\mu_i}$ are the
momentum transfer and the average of the momenta defined respectively
as $\Delta^{\mu_i} = p'^{\mu_i}-p^{\mu_i}$ and
$P^{\mu_i} =(p'^{\mu_i}+p^{\mu_i})/2$;  $u(p,\, s)$ and $\bar{u}(p',s')$
designate the spinor of the nucleon; $A_{n+1,i}(\Delta^2)$,
$B_{n+1,i}(\Delta^2)$, and $C_{n+1,0}(\Delta^2)$ represent the
generalized vector form factors (GVFFs) that are related to the Mellin
moments of the GPDs, which are given as  
\begin{widetext}
  \begin{eqnarray}
 \int_{-1}^1 x^n H(x,\xi,t) &=&
    \sum_{i=0,\mathrm{even}}^n (-2\xi)^i A_{n+1,i}(\Delta^2) +
\left.    (-2\xi)^{n+1} C_{n+1,0}
                    (\Delta^2)\right|_{n,\mathrm{odd}},\cr
 \int_{-1}^1 x^n E(x,\xi,t) &=&
    \sum_{i=0,\mathrm{even}}^n (-2\xi)^i B_{n+1,i}(\Delta^2) -
\left.    (-2\xi)^{n+1} C_{n+1,0}  (\Delta^2)\right|_{n,\mathrm{odd}}.     
  \end{eqnarray}
\end{widetext}
Here $H(x,\xi,t)$ and $E(x,\xi,t)$ are the twist-2 vector GPDs. The
usual Dirac and Pauli form factors are identified as the leading
GVFFs: 
\begin{equation}
F_1(\Delta^2) = A_{1,0} (\Delta^2),\;\;\;F_2=B_{1,0}(\Delta^2),  
\end{equation}
which are defined as the matrix element of the electromagnetic
current:  
\begin{eqnarray}
&&\left\langle N(p',s')\left|\bar{\psi}(0) \gamma_\mu \hat{Q} \psi(0)
  \right|N(p,s)\right\rangle  \cr
 &=&   
\bar{u}(p',s')\left[\gamma^{\mu}F_{1}
\left(\Delta^{2}\right)+i\frac{\sigma^{\mu\nu}\Delta_{\nu}}{2M_{N}}F_{2}
\left(\Delta^{2}\right)\right]u(p,s).
\end{eqnarray}

The nucleon matrix elements of the symmetric EMT operator are
parametrized in terms of the EMTFFs as
follows~\cite{Ji:1996ek,Polyakov:2002yz}:   
\begin{eqnarray}
\langle N(p^{\prime},s')|\hat{T}_{\mu\nu}(0)|N(p,s) \rangle  
&=&
\bar{u}(p^{\prime},\,
    s')\left[M_{2}(\Delta^2)\,\frac{P_{\mu}P_{\nu}}{M_{N}}\right. \cr
&&\hspace{-3cm} + J(\Delta^2)\ \frac{i(P_{\mu}\sigma_{\nu\rho}
+P_{\nu}\sigma_{\mu\rho}) \Delta^{\rho }}{2M_{N}}    \cr
&& \hspace{-3cm}\left. 
 +\,d_{1}(\Delta^2)\,\frac{\Delta_{\mu}
   \Delta_{\nu}-\Delta_{\mu\nu}\Delta^{2}}{5M_{N}}\right]u(p,\,  s). 
\label{Eq:EMTff} 
\end{eqnarray}
Similarly, the EMTFFs are related to the second moments of the vector
GPDs and in a such way  related to the GVFFs in the next-to-leading
order (NLO) as follows:  
\begin{eqnarray}
A_{2,0}\left(\Delta^2 \right) & = &
   \int_{-1}^{1}\mbox{d}x\,xH\left(x,0,\Delta^2
           \right)
   \;=\; M_{2}\left(\Delta^2 \right),\cr
B_{2,0}\left(\Delta^2\right)& = &
  \int_{-1}^{1}\mbox{d}x\,xE\left(x,0,t\right)
  =
  2J\left(\Delta^2\right)-M_{2}\left(\Delta^2
  \right),\cr   
C_{2,0}(\Delta^2) &=& \frac15 d_1(\Delta^2). 
\label{eq:gvffs}
\end{eqnarray}

In this work we want to examine the modification of the EMFFs and the
EMTFFs of the nucleon in nuclear medium. Let us first consider the EMFFs
of the nucleon. In the Breit frame one has $\Delta=\left(0,\,\bm
  \Delta \right)$ and $\bm p'=-\bm p$. Since it is more convenient to
introduce the positive definite square  
of the momentum transfer $Q^2 =-\Delta^2 >0$ to describe the form
factors, we will use it from now on. The Sachs EMFFs $G_{E}$ and
$G_{M}$ of the nucleon are expressed in terms of the Dirac and Pauli
form factors:   
\begin{eqnarray}
G_{E}\left(Q^{2}\right) & = &
F_{1}\left(Q^{2}\right)+\frac{Q^{2}}{4M_{N}^{2}}F_{2}
                                   \left(Q^{2}\right)\\   
G_{M}\left(Q^{2}\right) & = &
                              F_{1}\left(Q^{2}\right) +
                                   F_{2}\left(Q^{2}\right),  
\end{eqnarray}
which can be represented respectively as the Fourier transforms 
of the charge and current densities
\begin{eqnarray}
G_E(Q^2) &=&\int \mbox{d}^3r\, e^{i\bm \Delta \cdot \bm r}
                  J^0(r)\,,\cr  
G_M(Q^2) &=& m_N\int \mbox{d}^3re^{i  \bm \Delta \cdot \bm r}  
\left[\bm r \times {\bm J}(\bm r)\right]_3.
\end{eqnarray} 
Here $J^0$ and $\bm J$ denote respectively the charge and current
densities. Note that the EM current $J^\mu$ is defined as the sum of
the baryonic current $B^\mu$ and the third component of the isovector
current $\bm V^\mu$.  The final expressions for the in-medium modified
isoscalar and isovector FFs are derived as
\begin{widetext}
\begin{eqnarray}
G_{E}^{S}\left(Q^{2}\right) & = & -\frac{m_{\omega}^{*2}}{3\sqrt{\zeta}g}
\int_{0}^{\infty}r^{2}j_{0}\left(Qr\right)\omega(r)\mbox{d}r\,,
\label{EMffbegin}
\\
G_{M}^{S}\left(Q^{2}\right) & = & -\frac{m_{\omega}^{*2}}{3\sqrt{\zeta}g}
\frac{M_{N}}{\lambda^{*}}2\pi\int_{0}^{\infty}r^{2}\frac{j_{1}\left(Qr\right)}
{Qr}\phi(r) \mbox{d}r\,,\\
G_{E}^{V}\left(Q^{2}\right) & = & \frac{4\pi}{\lambda^{*}}
\int_{0}^{\infty}j_{0}\left(Qr\right)\left[\frac{f_{\pi}r^{2}}{3}\left\{ 
4\sin^{4}\frac{F}{2}+\left(1+2\cos F\right)\xi_{1}+\xi_{2}\right\} 
+\frac{g\sqrt{\zeta}}{8\pi^{2}}\phi F'\sin^{2}F\right]\mbox{d}r\,,\\
G_{M}^{V}\left(Q^{2}\right) & = & \frac{8\pi}{3}M_{N}\int_{0}^{\infty}r^{2}
\frac{j_{1}\left(Qr\right)}{Qr}\left[2f_{\pi}^{2}\left(2\sin^{4}\frac{F}{2}
-2\left(1-\alpha_{p}\right)\frac{1}{4}\sin^{2}F-G\cos F\right)
+\frac{3g\sqrt{\zeta}}{4\pi^{2}}\omega F'\sin^{2}F\right]\,,
\label{EMffend}
\end{eqnarray}
\end{widetext}
where the detailed expressions for the profile functions, $\omega(r)$,
$\phi(r)$, $F(r)$, $\xi_1(r)$, and $\xi_2(r)$, can be found in
Ref.~\cite{Jung:2012sy}.  
 The proton and neutron EMFFs are expressed in terms of the isoscalar
 and isovector FFs  
\begin{equation}
G_{E,M}^{p,n}(Q^2) \;=\; G_{E,M}^{S}(Q^2) + \tau_3
G_{E,M}^{V}(Q^2), 
\label{eq:isoff}     
\end{equation}
where $\tau_3$ is the eigenvalue of $\hat{\tau}_3$ for a given nucleon 
isospin state. At the zero momentum transfer ($Q^2 = 0$) the EMFFs are
normalized as  
\begin{eqnarray}
G_{E}^{p}\left(0\right)=1,\quad
G_{E}^{n}\left(0\right)=0,\quad G_{M}^{p,n}\left(0\right)=\mu_{p,n}.
\label{EMFF}
\end{eqnarray}
Since we already have studied the EMTFFs in nuclear matter within the
present  approach~\cite{Jung:2014jja}, we  refer to
Ref.~\cite{Jung:2014jja} for details.  

Once we obtain the EMFFs and the EMTFFs of the nucleon, we can proceed
to derive the transverse quark charge densities inside a nucleon,
which show how the charges and magnetizations of the quarks are
distributed in the transverse plane inside a
nucleon~\cite{Burkardt:2002hr,Kelly:2002if}. The transverse charge
density inside an unpolarized nucleon is defined as the
two-dimensional Fourier transform of the Dirac form factor: 
\begin{widetext}
\begin{eqnarray}
\rho_{\rm ch}&=& \frac{1}{(2\pi)^ 2} \int d^2 q e^{i{\bm q}\cdot{\bm b}}F_1 (Q^2)  
=\int_0^\infty \frac{dQ}{2\pi}\, Q J_0(Qb) F_1(Q^2) =   
\int_0^\infty \frac{dQ}{2\pi}\, Q J_0(Qb) \frac{G_E(Q^2)
  +\tau\,G_M(Q^2)}{1+\tau}, 
\label{eq:rhoch}
\end{eqnarray}
\end{widetext}
where $b$ designates the impact parameter, i.e. the distance in the
transverse plane to the place where the density is being probed, and
$J_0$ denotes the Bessel function of order
zero~\cite{Miller:2007uy,Miller:2010nz}.   
The anomalous magnetisation density in the transverse
plane~\cite{Miller:2010nz,Venkat:2010by} is defined as 
\begin{equation}
\rho_{\rm m} \;=\; -b\frac{d\;}{db}\rho_2(b)  \;=\; b \,\int_0^\infty
\frac{dQ}{2\pi}\, Q^2 J_1(Qb) F_2(Q^2),  
\label{eq:rhoM}  
\end{equation}
where $\rho_2(b)$ is directly given by the two-dimensional Fourier
transform of the Pauli form factor: 
\begin{equation}
\rho_2(b)=\int_0^\infty \frac{dQ}{2\pi}\, Q J_0(Qb) F_2(Q^2).
\label{eq:rho2}
\end{equation}

Assume that the nucleon is transversely polarized along the $x$
axis. Then the polarization of the nucleon can be expressed in terms
of the transverse spin operator of the nucleon $\bm S_\perp =
\cos\varphi_S \hat{e}_x +\sin\varphi_S\hat{e}_y$, so that the
transverse charge density inside a transversely polarized nucleon is
written as~\cite{Carlson:2007xd}   
\begin{equation}
  \label{eq:poltrans}
\rho_T(\bm b) = \rho_{\mathrm{ch}} -\sin(\varphi_b-\varphi_S)
\frac1{2M_N} \rho_{\mathrm{m}} (b),   
\end{equation}
where the angle $\varphi_b$ is defined in the position vector $\bm b$
that stands for the impact parameter or the transverse distance from
the center of the nucleon in the transverse plane $\bm
b=b(\cos\varphi_b \hat{\bm e}_x + \sin \varphi_b \hat{\bm e}_y)$. 

Since the EMTFFs are identified as the generalized vector FFs in the
isocalar channel, one can also define the transverse isoscalar
densities in the case of the EMTFFs, which takes the following form: 
\begin{eqnarray}
\rho_{20}\left(b\right)
&=&\int_{0}^{\infty}\frac{\mbox{d}Q}{2\pi}\,Q J_{0}
\left(Qb\right)A_{2,0}\left(Q^{2}\right)\,.
\end{eqnarray}
When the nucleon is polarized along the $x$ axis in the transverse
plane, the transverse isoscalar density inside the polarized nucleon
is defined as 
\begin{eqnarray}
\rho_{20,T}(\boldsymbol{b})&=&	\rho_{20}\left(b\right)
-\sin\left(\phi_{b}-\phi_{S}\right)\cr
&&\times 
\int_{0}^{\infty}\frac{Q^{2}\mbox{d}Q}{4\pi M_{N}}J_{1}\left(Qb\right)
B_{2,0}\left(Q^2\right)\,.
\end{eqnarray}

\section{Results and discussions}

In this Section we present the numerical results of the form factors
and related observables and discuss their physical implications. 
\subsection{Electromagnetic form factors and Transverse Charge
  densities} 
We first show the results for the traditional charge and magnetization
radii and the magnetic moments of the proton and the neutron. 
Table~\ref{tab1} lists them in free space calculated within two
different models. It is already well known that the
$\pi$-$\rho$-$\omega$ soliton model overestimates the magnetic
moments of the nucleon. On the other hand, the results of the
traditional charge and magnetization radii of the proton are in good
agreement with the experimental data. Note that there is almost no
difference between model I and model II in free space, as expected.  
\begin{table}[bt]
\begin{ruledtabular}
\begin{tabular}{cc|c|c|c}
&&Model I& Model II &Experiment \\
\colrule
$\left\langle r_{E}^{2}\right\rangle _{p}^{1/2}$ 
&{$\left[\mbox{fm}\right]$}
 & 0.93 & 0.93 &  0.86 \\
$\left\langle r_{M}^{2}\right\rangle _{p}^{1/2}$   
&{$\left[\mbox{fm}\right]$}
& 0.87 & 0.87 &  0.78 \\
$\left\langle r_{E}^{2}\right\rangle _{n}$ 
&{$\left[\mbox{fm}^{2}\right]$}
& -0.23 & -0.23 & -0.12 \\
$\left\langle r_{M}^{2}\right\rangle _{n}^{1/2}$ 
&{$\left[\mbox{fm}\right]$}
& 0.88 & 0.88 & 0.86 \\
$\mu_{p}$ &{$\left[\mu_N\right]$}
& 3.37  & 3.39 &  2.79 \\
$\mu_{n}$ & {$\left[\mu_N \right]$}
& -2.58 & -2.61 & -1.91  \\
 $\left|\mu_{p}/\mu_{n}\right|$ &&
 1.31 & 1.30 & 1.46
\end{tabular}
\end{ruledtabular}
\caption{The electromagnetic properties of the nucleons in free space.  
The magnetic moments of the proton and the neutron are given in
the unit of the nuclear magneton ($\mu_N$).}
\label{tab1}
\end{table}

If the nucleon is embedded into nuclear medium, then its properties
undergo the changes due to the interaction with the surrounding
environment. The results listed in Table~\ref{tab2} demonstrate
possible medium modifications of the EM radii and the 
magnetic moments of the nucleons at normal nuclear matter density. 
\begin{table}[hbt]
\begin{ruledtabular}
\begin{tabular}{cc|c|c}
&&Model I& Model II \\
\colrule
$\left\langle r_{E}^{*2}\right\rangle _{p}^{1/2}$ 
&{$\left[\mbox{fm}\right]$}
 & 1.17 & 1.08 \\
$\left\langle r_{M}^{*2}\right\rangle _{p}^{1/2}$ 
&{$\left[\mbox{fm}\right]$}
 & 1.17 & 1.14 \\
$\left\langle r_{E}^{*2}\right\rangle _{n}$ 
&{$\left[\mbox{fm}^{2}\right]$}
& -0.22 & -0.40 \\
$\left\langle r_{M}^{*2}\right\rangle _{n}^{1/2}$ 
&{$\left[\mbox{fm}\right]$}
 & 1.18 & 1.17 \\
$\mu_{p}^*$ &{$\left[\mu_N \right]$}
& 5.23  & 5.41 \\
$\mu_{n}^*$ & {$\left[\mu_N \right]$}
& -4.56 & -4.73 \\
 $\left|\mu_{p}^*/\mu_{n}^*\right|$ &&
 1.15 & 1.14
\end{tabular}
\end{ruledtabular}
\caption{The electromagnetic properties of the nucleons in nuclear
  medium at normal nuclear matter density $\rho_0$. } 
\label{tab2}
\end{table}
The size of the proton charge radius in medium turns out to be larger
than that in free space. Both the results from Model I and Model II 
show similar tendencies. It indicates that the nucleon tends to buldge
out in nuclear medium. On the other hand, Model I and Model II yield
different results. While the neutron charge radius from Model I
is almost the same as that in free space, its magnitude from Model II
is drastically increased. In fact, the neutron radius is a rather
difficult observable to describe theoretically because it comes from
the subtraction between the isoscalar and isovector FFs (see
Eq.(\ref{eq:isoff})). As will be discussed later, the traditional
charge density of the neutron is very different from the transverse
charge density. In addition, the medium effects affect strongly the
radial dependence of the neutron charge distribution in comparison
with the proton one. Thus, it is difficult to draw any conclusion
about the changes of the neutron size in medium, based on the
traditional neutron charge density. However, we will soon see that the
transverse charge density inside a nucleon will clearly show that both
the proton and the neutron swell in nuclear medium.  
\begin{figure*}[ht]
\includegraphics[scale=0.4]{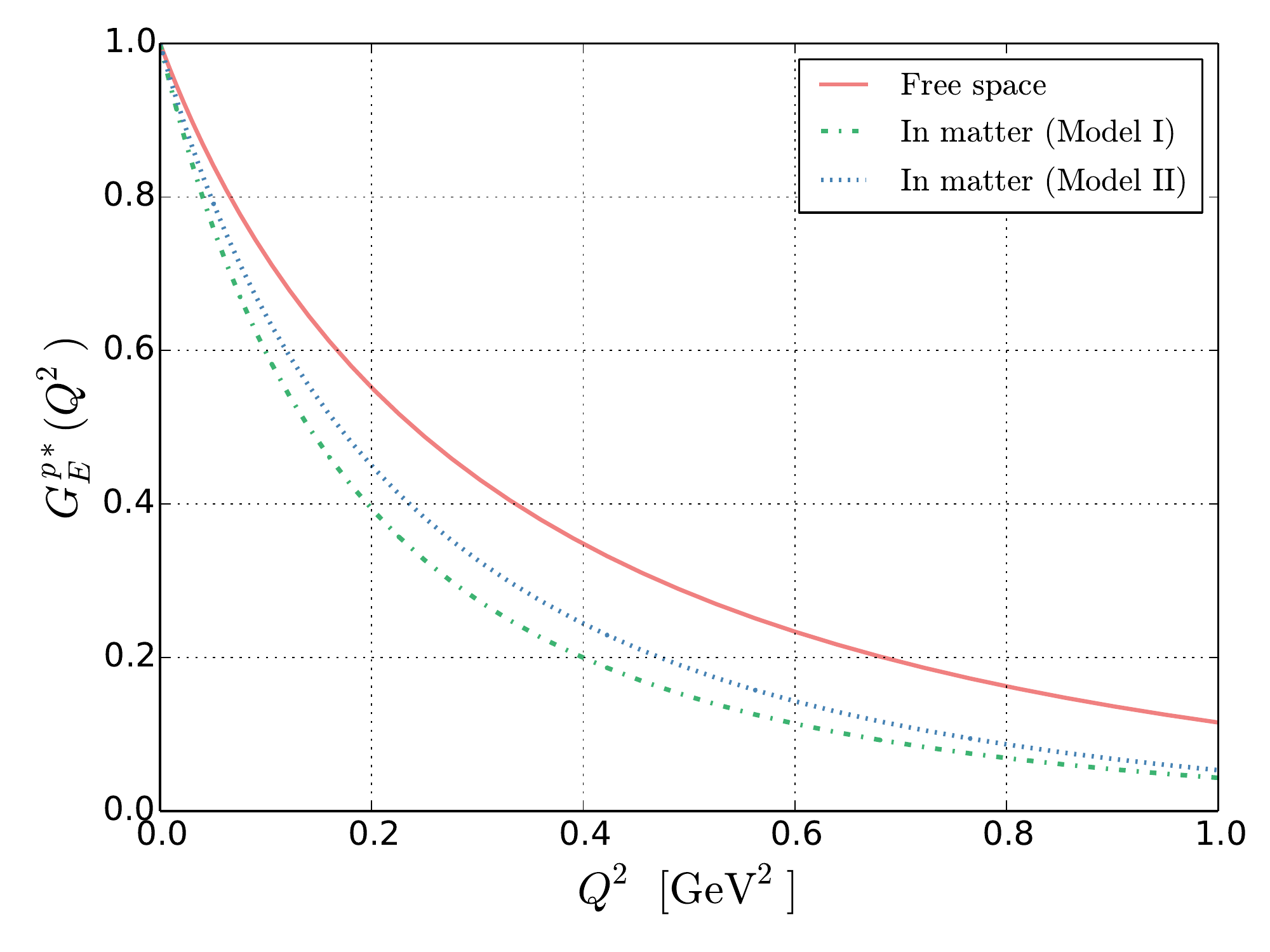}\includegraphics[scale=0.4]{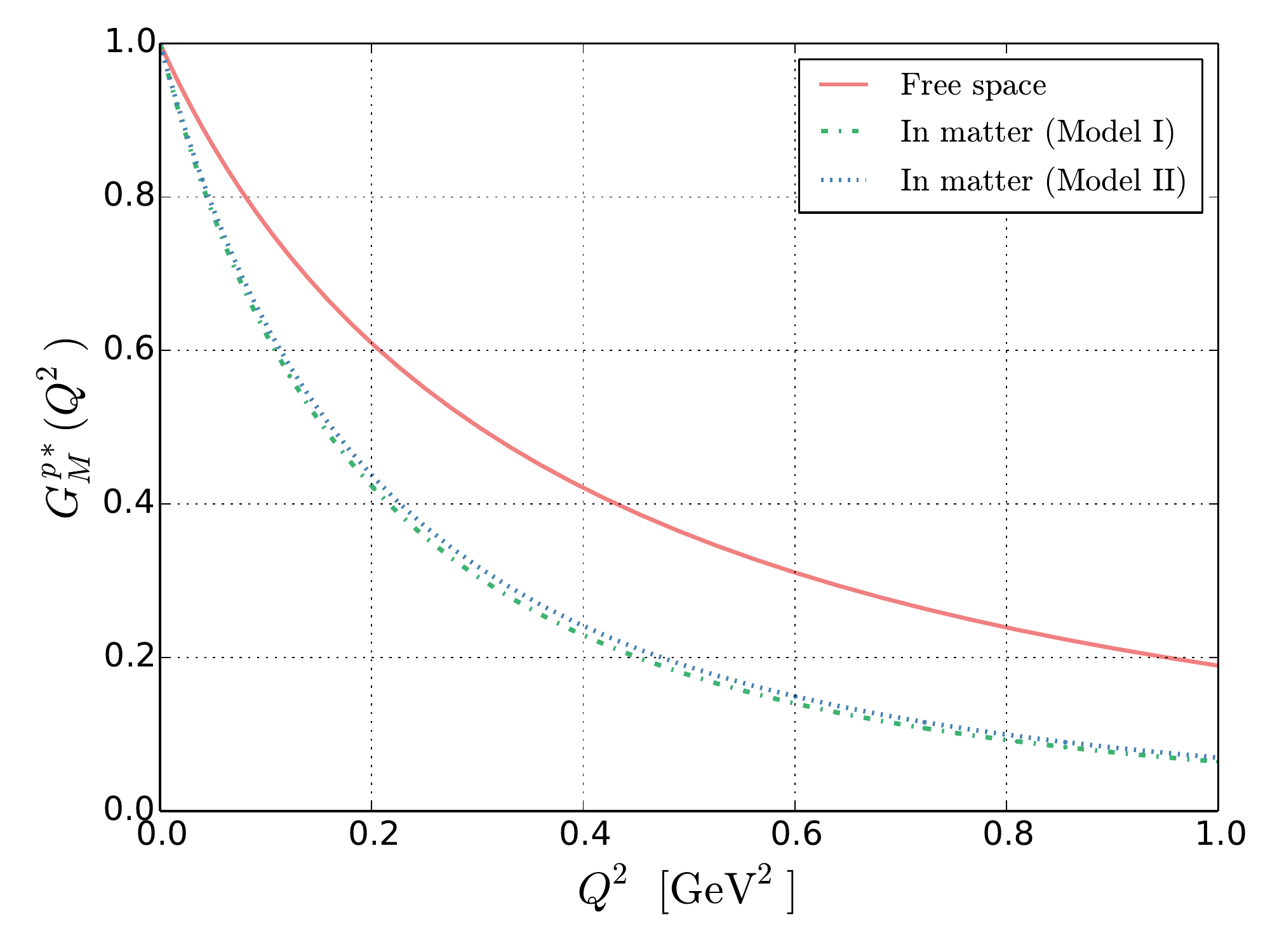}\\
\includegraphics[scale=0.4]{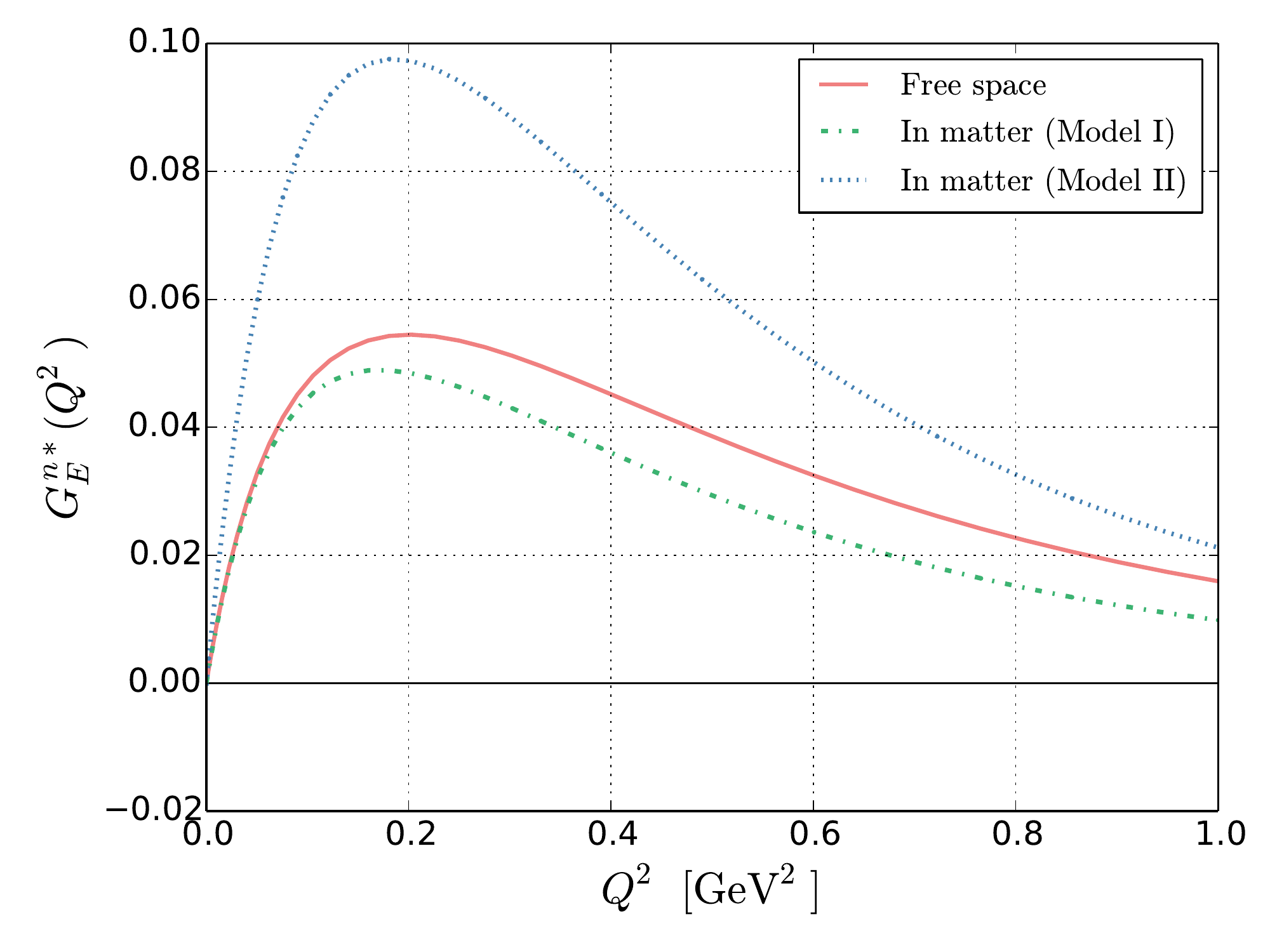}\includegraphics[scale=0.4]{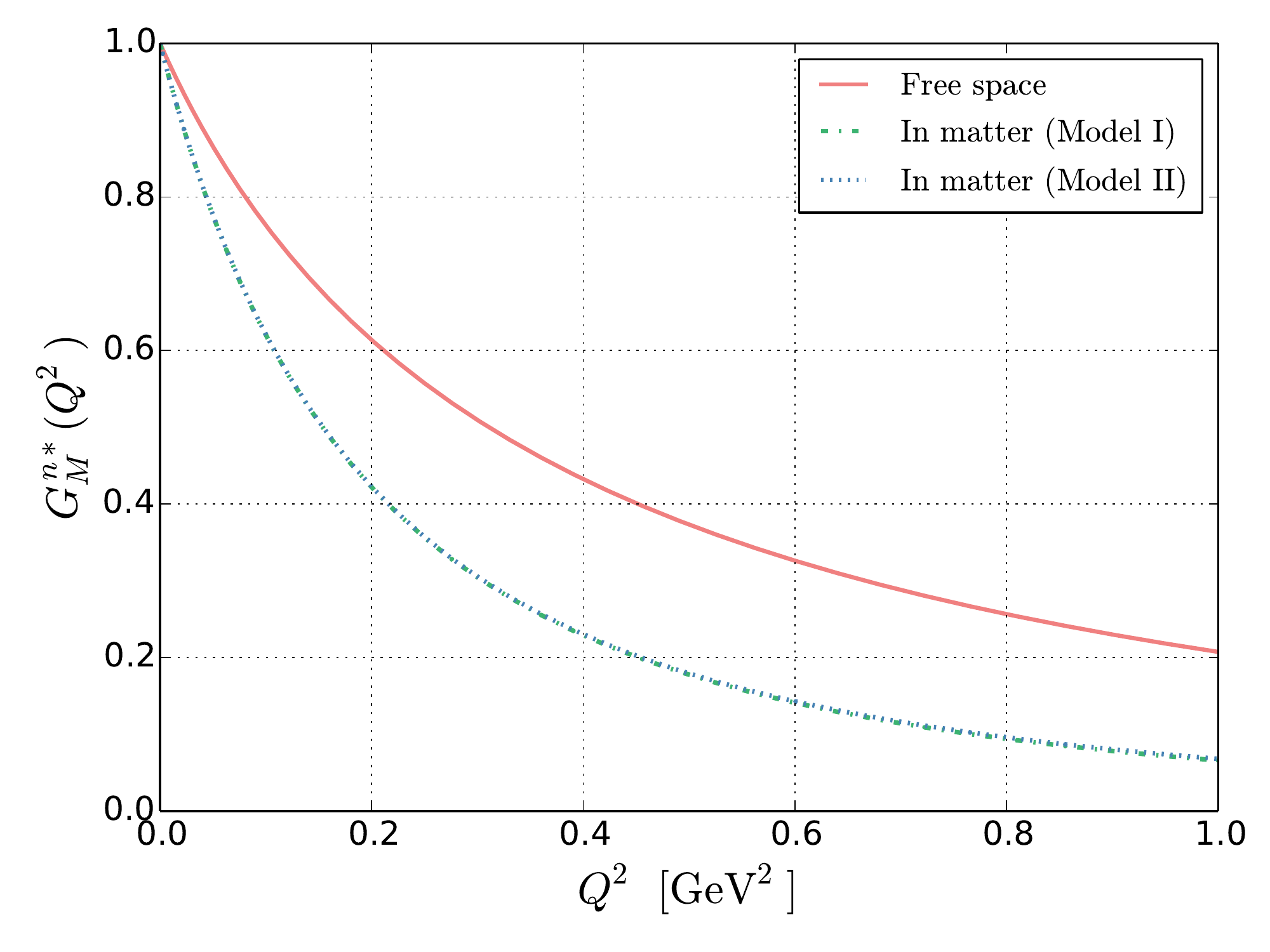}
\caption{The electric and magnetic form factors of the proton are
  drawn respectively in the upper-left and upper-right panels, and
  those of the neutron are depicted in the lower panels in the same
  manner as functions of $Q^2$. The solid curve represents the form
  factors in free space, while the dotted and dotted-dashed ones
  designate, respectively, those from 
Model I and Model~II in nuclear matter.} 
\label{fig:EMFF}
\end{figure*}
The magnitudes of the magnetic moments of both the proton and the
neutron become quite larger in nuclear medium than in free space by
approximately $40\,\%$, as shown in Table~\ref{tab2}.
The medium effects turn out to be even larger on the neutron magnetic
moment than the proton one as observed in the results of their
ratio $\left|\mu_{p}^*/\mu_{n}^*\right|$. The reason can be found in
the fact that the magnetization density becomes broadened in
medium, which will be shown soon. Since the operator of the magnetic
moment is proportional to the distance from the center of the nucleon,
the nucleon magnetic moments in general tend to increase in nuclear
medium. This also indicates indirectly that the nucleon swells in
nuclear matter.  

Figure~\ref{fig:EMFF} depicts the results for the EMFFs of the
proton and the neutron as functions of $Q^2$.
As was expected from the charge and magnetic radii of the proton, the
EMFFs of the proton in medium fall off
faster than those in free space as $Q^2$ increases. The general
tendency of the form factors remains almost unchanged in the case of
both Model I and Model II. When it comes to the electric FF of the
neutron, however, the result from Model~I is very different 
from that obtained from Model II. We already have seen that Model I
and Model II give rather different results for the neutron charge 
radii. The difference arises from the fact that the $\omega$ meson is
treated in a distinctive way. In Model I, both the $\rho$ and $\omega$
mesons are treated on an equal footing. That is, both the vector
mesons undergo changes in the same manner. On the other hand, the
$\omega$ meson is kept to be same as in free space in Model~II. Since
the proton electric FF is given as the sum of the isoscalar and
isovector form factors as shown in Eq.(\ref{eq:isoff}), the difference
between Model~I and Model~II is marginal (see the results for the
electric FFs of the proton in Fig.~\ref{fig:EMFF}). However, the
neutron electric FF comes from the subtraction of the isovector FF
from the isoscalar one. Considering the fact that the $\rho$ meson
contributes only to the isovector FF whereas the $\omega$ meson comes
into play only in the isoscalar FF, we can easily see that the changes
of both the $\rho$ and $\omega$ mesons are more or less compensated in
Model~I. However, in Model~II, the isoscalar FF remains intact while
the isovector FF is modified, which leads to the amplication of the
electric FF of the nucleon (see the lower panel of Fig.~\ref{fig:EMFF}). 
It is interesting to note that the results for the neutron from Model
I is very similar to those from
Ref.~\cite{Yakhshiev:2012zz}. Considering the fact that the Skyrme
term in Ref.~\cite{Yakhshiev:2012zz} is related to the vector mesons
by the resonance saturation~\cite{Ecker:1988te}, The characteristics
of Model I is closer to the medium-modified Skyrme model in which both
the pion kinetic and Skyrme terms are modified, compared to Model II.

\begin{figure*}
\includegraphics[scale=0.4]{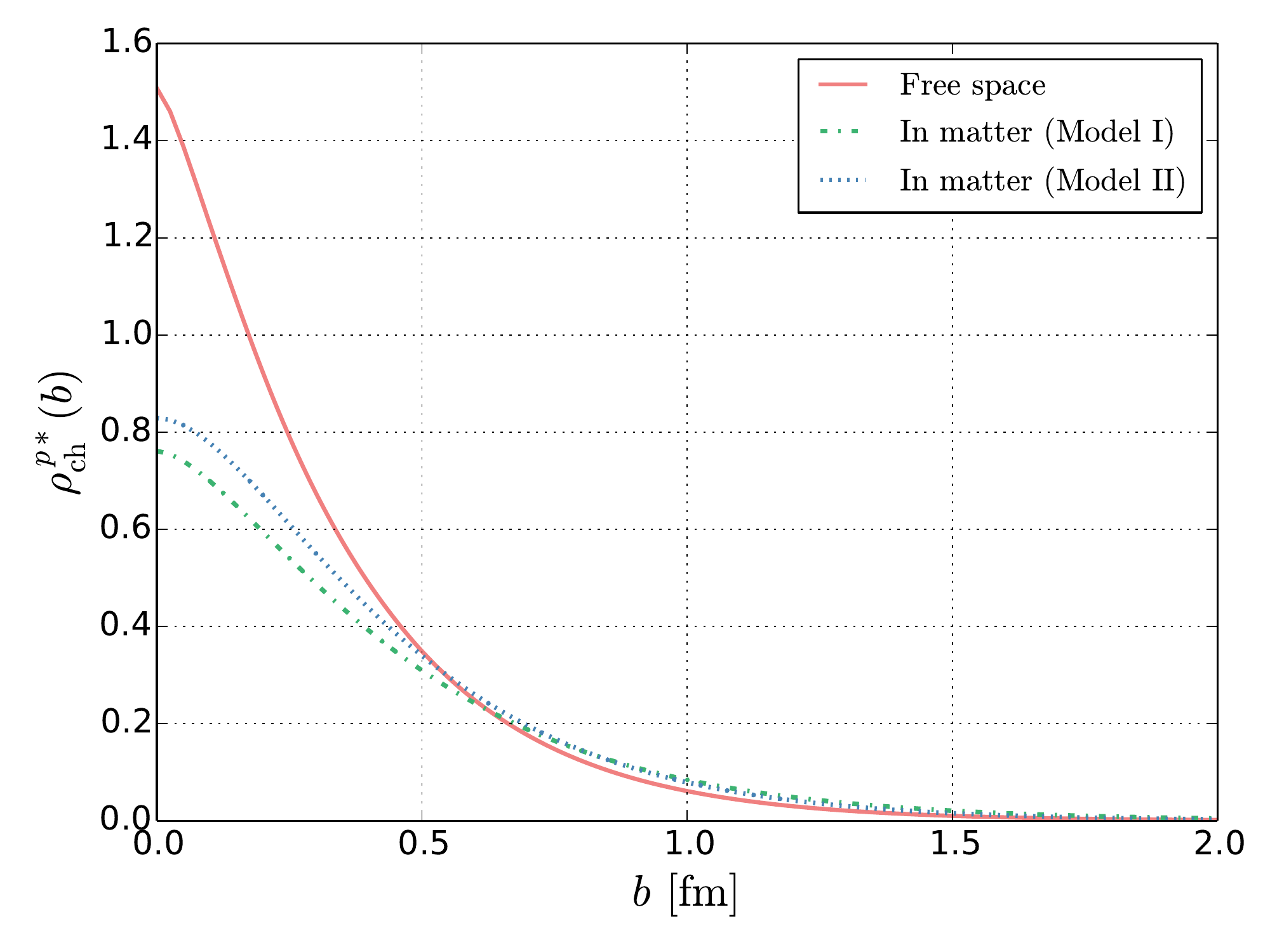}\includegraphics[scale=0.4]{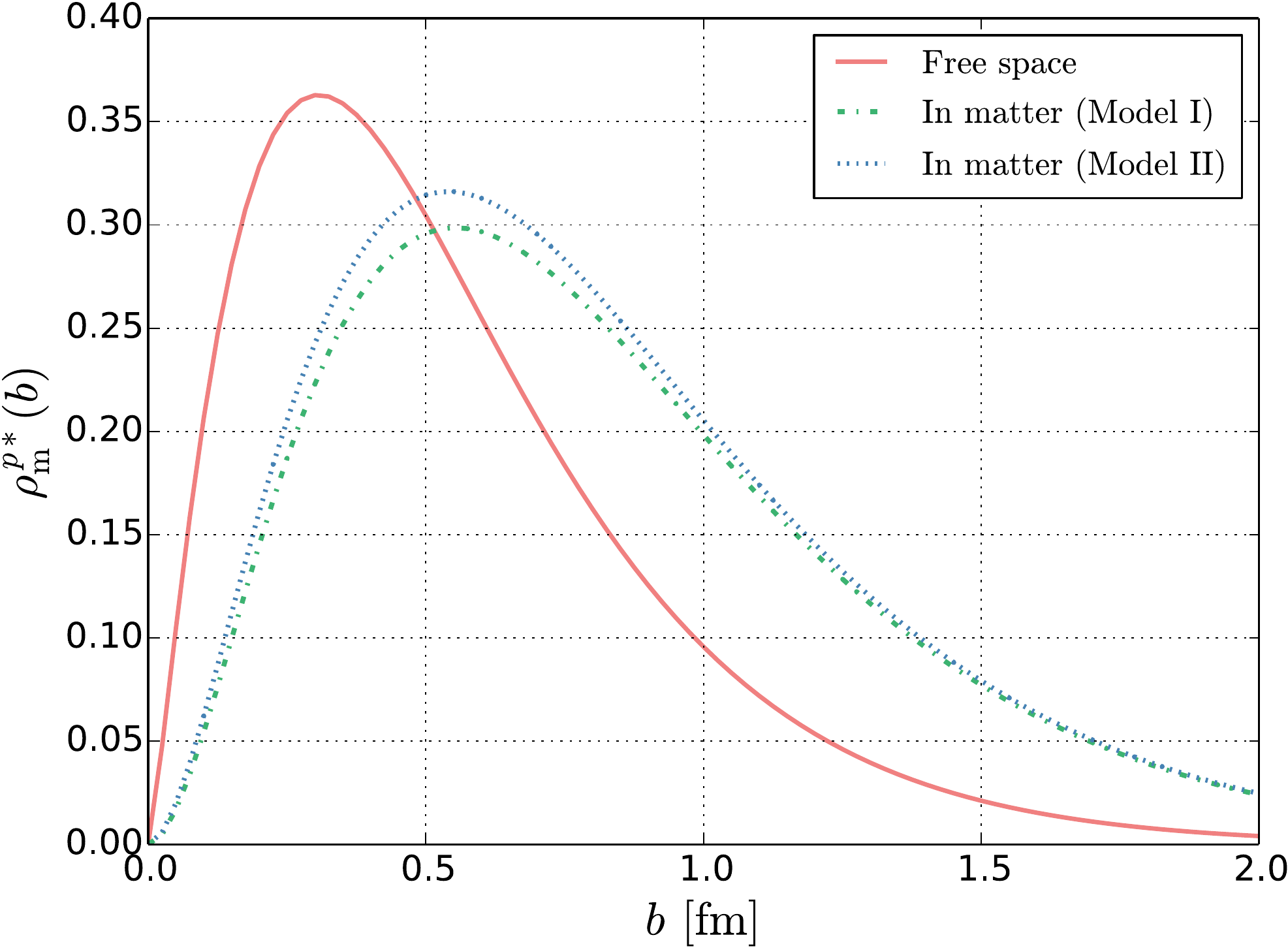}\\
\includegraphics[scale=0.4]{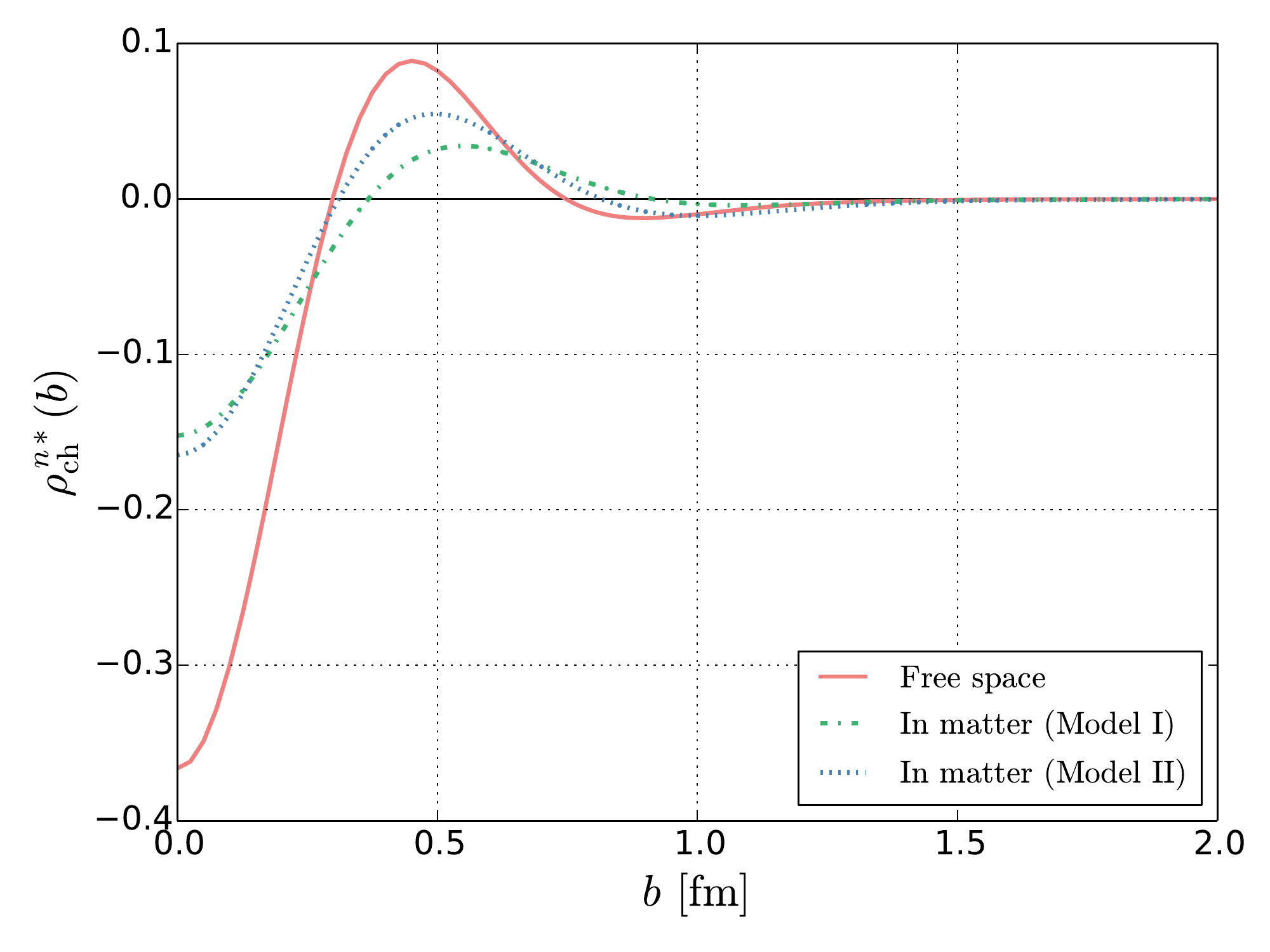}\includegraphics[scale=0.4]{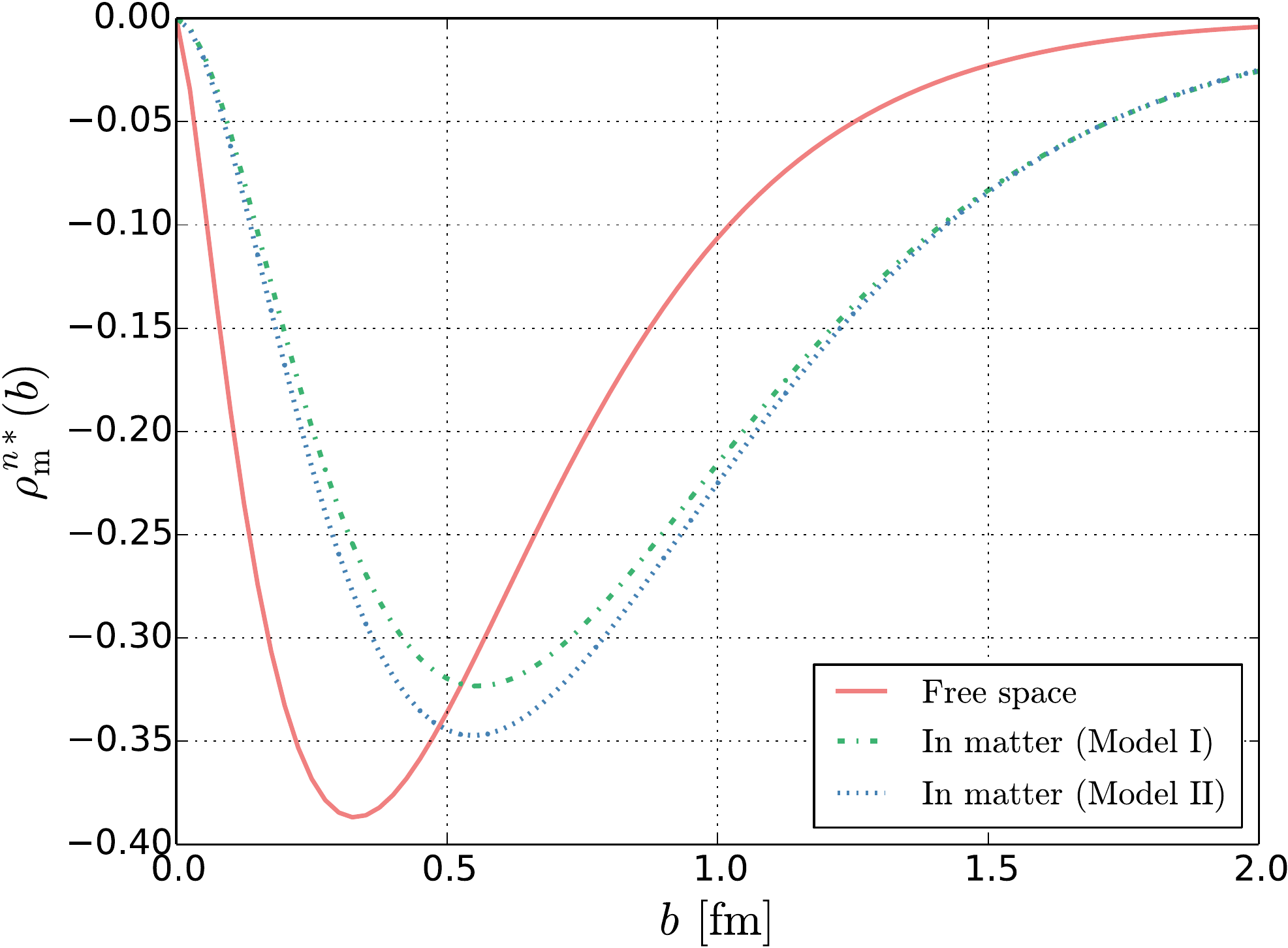}
\caption{The transverse charge densities inside an unpolarized proton
  with $b_x = 0$ are plotted in the upper-left and the
  upper-right panels, respectively, and those inside a neutron are
  depicted in the lower panels in the same manner as functions of
  the impact parameter $b$.   The solid curve represents the
  transverse densities in free space, while the dotted and dotted-dashed ones
  designate, respectively, those from Model I and Model II in nuclear
  matter.}  
\label{perpEMchargeDist}
\end{figure*}
The results for the transverse charge and magnetization distributions
inside an unpolarized proton are drawn in the upper-left and
upper-right panels, respectively, with $b_x=0$. The medium-modified
transverse charge densities near the center of the proton are reduced
drastically but get larger as $b$ increases. It indicates that the
transverse size of the nucleon becomes larger in nuclear medium.
As for the transverse magnetization densities, we find that the 
densties in medium are shifted and broadened in comparison with that
in free space. It also implies that the in-medium nucleon swells
relatively to the free space one.  

It is already well known that the transverse charge density inside an
unpolarized neutron provides a new aspect on the structure of the
neutron~\cite{Miller:2007uy,Miller:2010nz}. Considering the fact that
the transverse charge density inside a nucleon has a physical meaning
of the probablility of finding a quark inside a nucleon, we can see
from the results for the transverse charge densities inside a neutron,
which are depicted in the lower-left panel of
Fig.~\ref{perpEMchargeDist}, that the negative charged quarks,
i.e. down quarks are more probably found in the vicinity of the center
of the neutron whereas the positive charged quarks or up quarks are
located in outer regions inside a neutron. This is very much different
from the usual and traditional understanding of the neutron charge
distribution in which the positive charge is found near the center of
the neutron while the negative charge is placed in outer regions. In
nuclear matter,  the transverse charge density inside
a neutron has the same tendency but the magnitude of the densities is
reduced and is broadened, as shown in the lower-left panel of 
Fig.~\ref{perpEMchargeDist}. It imlpies that the size of the neutron
is also extended in nuclear medium. The transverse magnetization
density inside a neutron is similarily modified in nuclear medium as
that inside a proton.  

\begin{table}[hbt]
\begin{ruledtabular}
\begin{tabular}{ll|c|c|c}
&& Free Space&Model I& Model II\\
\colrule
{$\left\langle b_{\mathrm{ch}}^{2}\right\rangle^{1/2}_p$} 
&{$\left[\mbox{fm}\right]$}
& 0.70 & 0.90 &  0.81   \\
{$\left\langle b_{\mathrm{m}}^{2}\right\rangle^{1/2}_p$} 
&{$\left[\mbox{fm}\right]$}
& 0.89  & 1.65  &  1.67   \\
{$\left\langle b_{\mathrm{ch}}^{2}\right\rangle_n$} 
&{$\left[\mbox{fm}^{2}\right]$}
& -0.023 & -0.015 & -0.042   \\
{$\left\langle b_{\mathrm{m}}^{2}\right\rangle_n$} 
&{$\left[\mbox{fm}^{2}\right]$}
& -0.85   & -2.89  &  -2.89   \\
\end{tabular}
\end{ruledtabular}
\caption{The transverse charge and magnetization radii of the
  unpolarized proton and the neutron. The results in free
  space and in nuclear matter at normal nuclear matter density,
  $\rho_0$, are presented.}
\label{tab3}
\end{table}
To see the swelling of the nucleon in nuclear matter more clearly, we 
define the transverse mean square charge 
and magnetization radii of the nucleon as follows 
\begin{eqnarray}
\left\langle b_{\mathrm{ch,\,m}}^{2}\right\rangle_{p,n}
=\int {\rm d}^{2}b\,b^{2}\rho_{\mathrm{ch,\,m}}^{p,n} \left(b\right),
\end{eqnarray}
where the transverse charge density, $\rho_{\mathrm{ch}}$, and the
transverse magnetization density, $\rho_{\mathrm{m}}$ are defined in
Eq.~(\ref{eq:rhoch}) and Eq.~(\ref{eq:rhoM}), respectively.  
The results for the transverse charge and magnetization radii of the
proton and the neutron are listed in Table~\ref{tab3}. The transverse
mean square charge radius of the proton in nuclear medium is
increased approximately by $20\,\%$. On the other hand, the
medium-modified trnsverse mean magnetization radius of the proton
becomes almost twice as large as that in free space. 
In the case of the neutron, the result from Model I shows slightly
smaller than that in free space whereas the result from Model II is
almost about two times larger than that in free space. As we have
already discussed previously, the role of the $\omega$ meson becomes
much more influential in the case of the neutron than in the proton
case.  

\begin{figure*}
\begin{centering}
\includegraphics[scale=0.55]{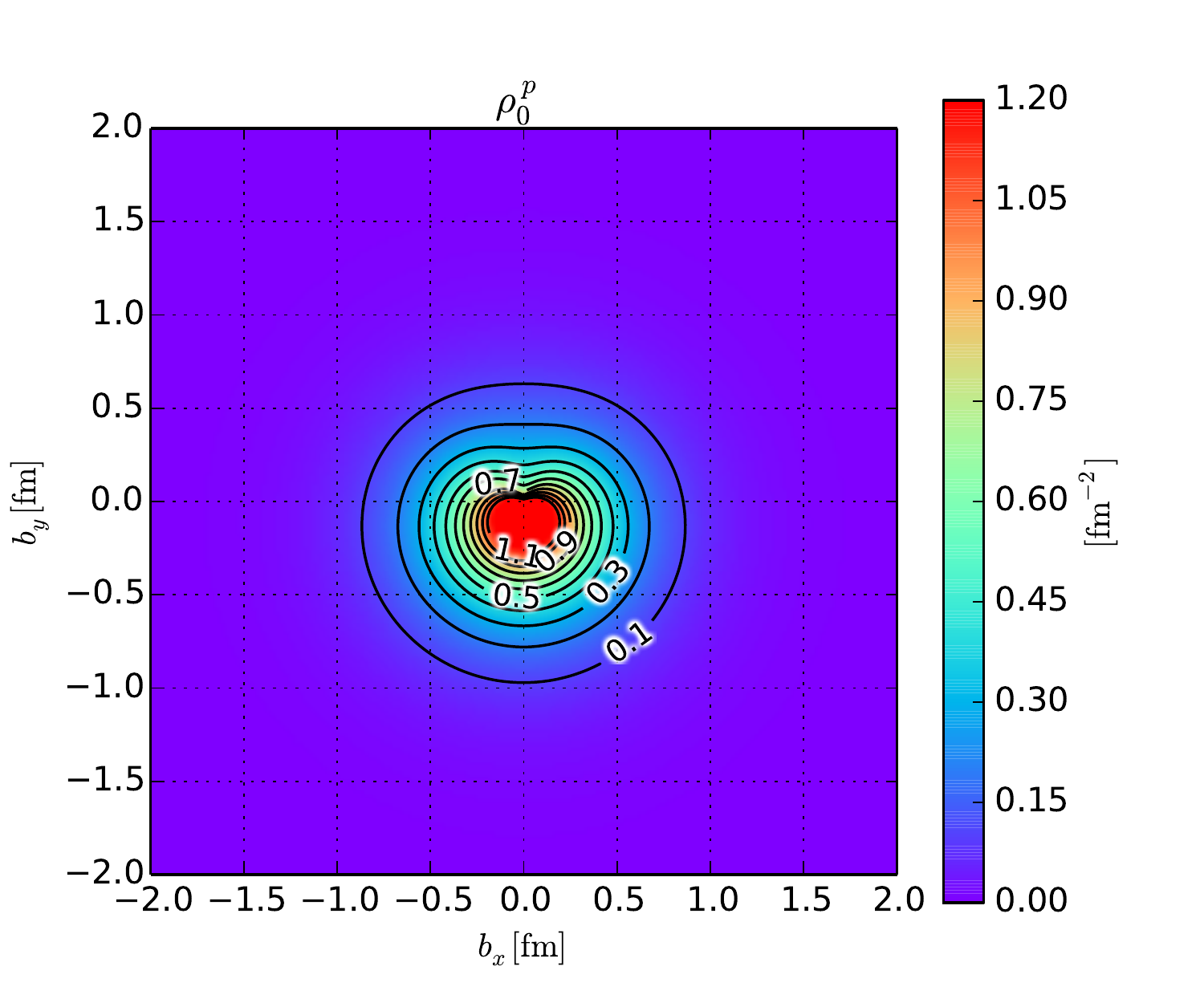}
\includegraphics[scale=0.55]{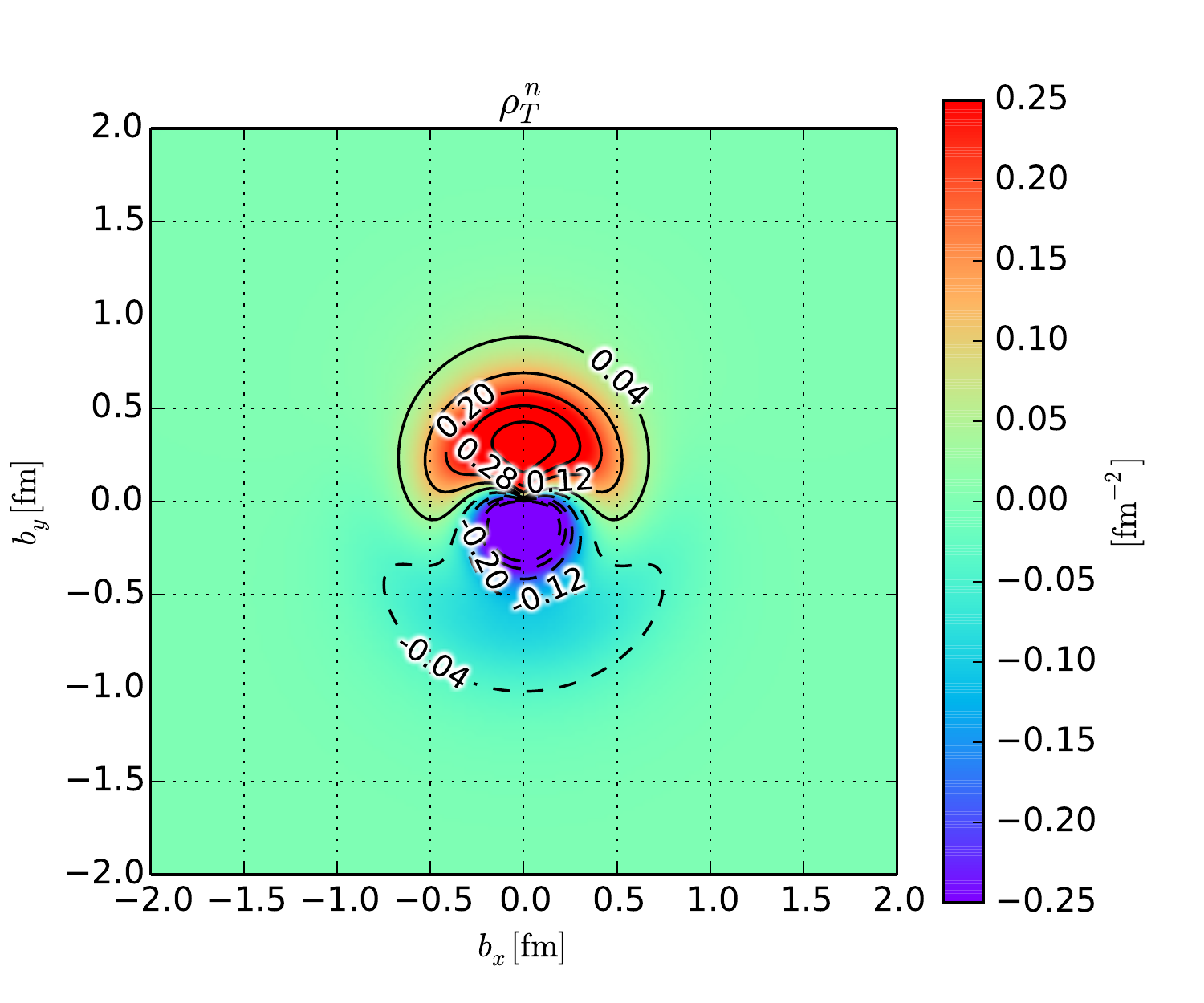}
\par\end{centering}
\caption{Transverse charge densities inside the polarized proton (left panel) 
and neutron (right panel) in free space.}
\label{fig:polNPTCfs}
\end{figure*}

\begin{figure*}
\begin{centering}
\includegraphics[scale=0.55]{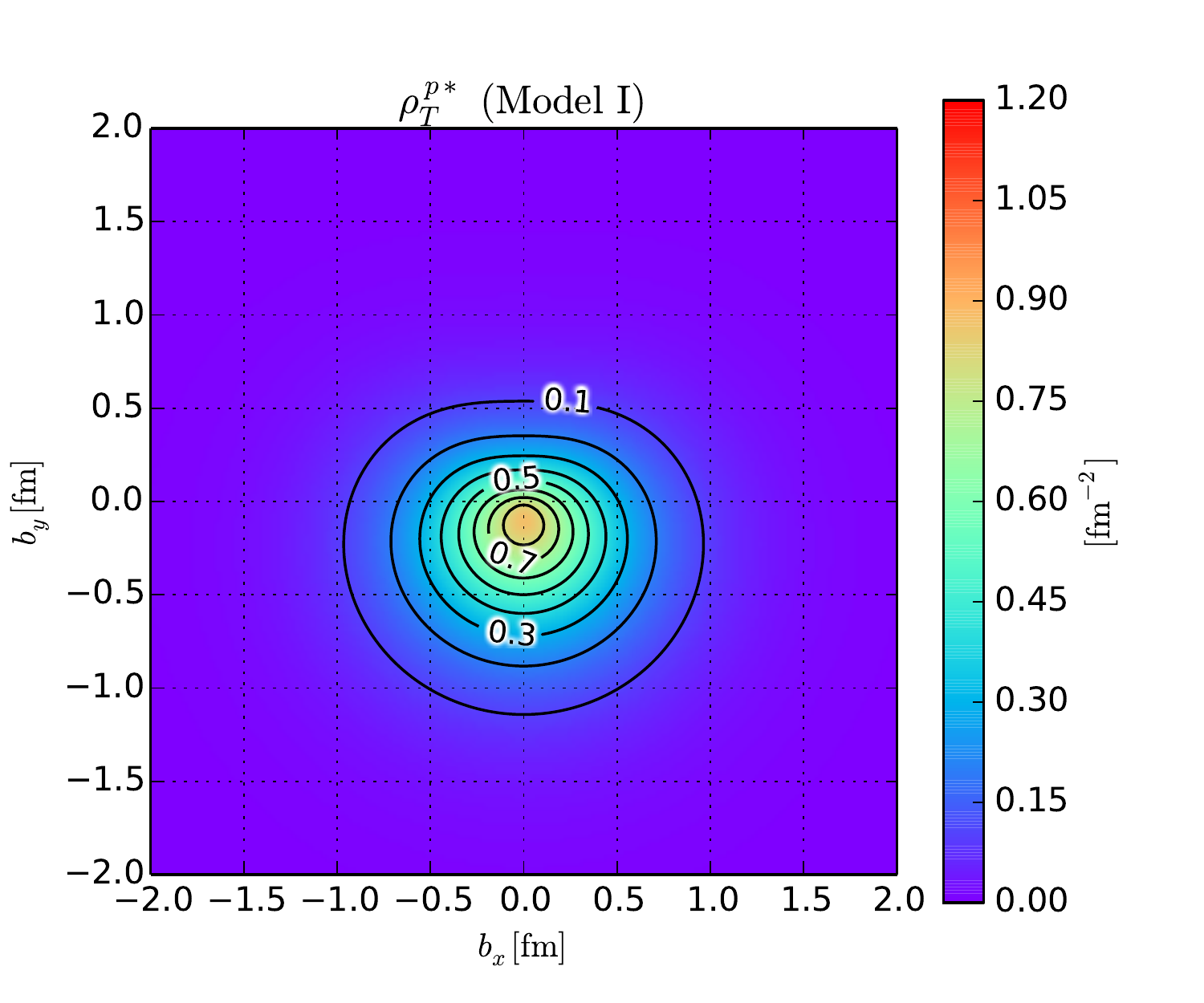}
\includegraphics[scale=0.55]{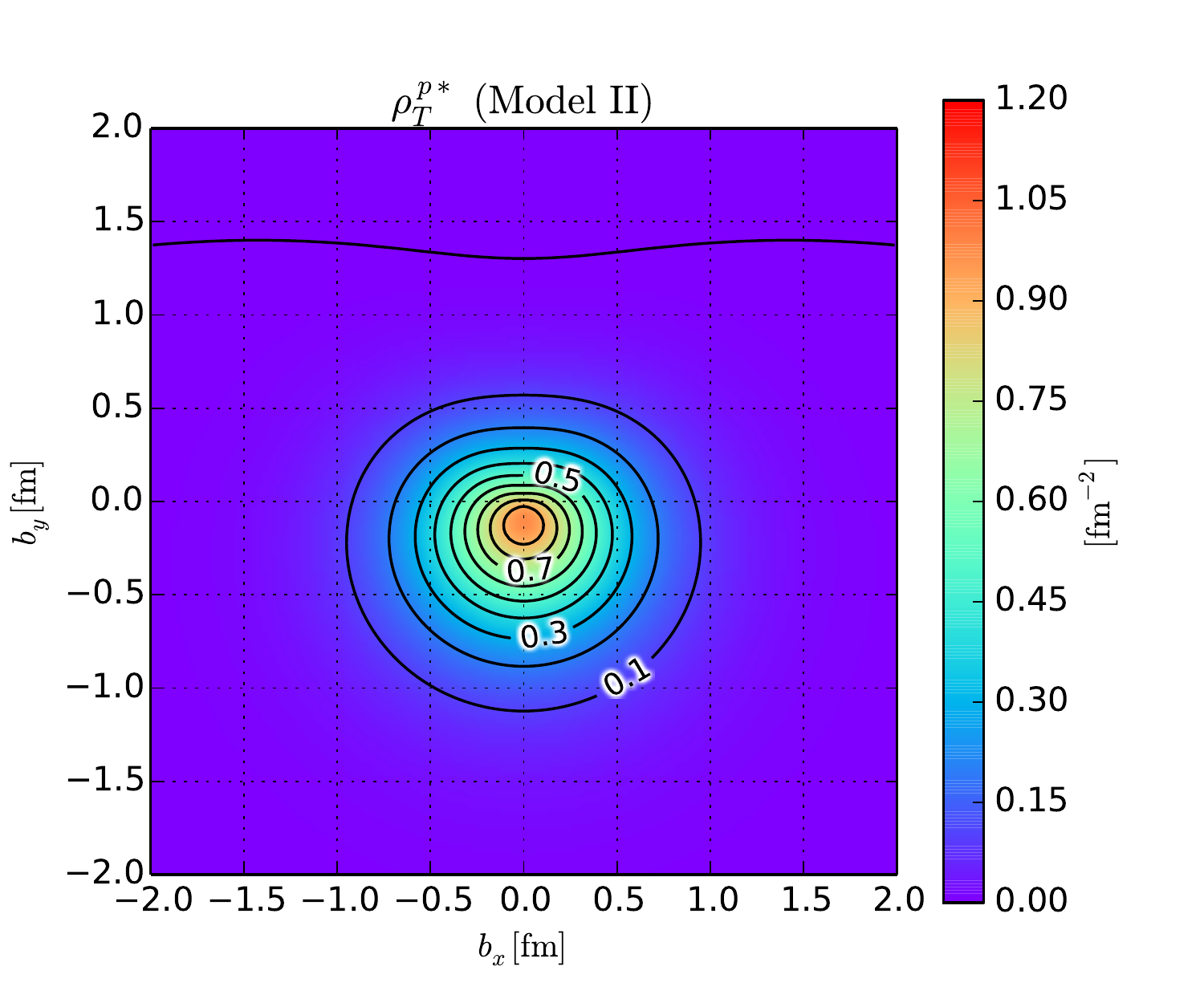}\\
\includegraphics[scale=0.55]{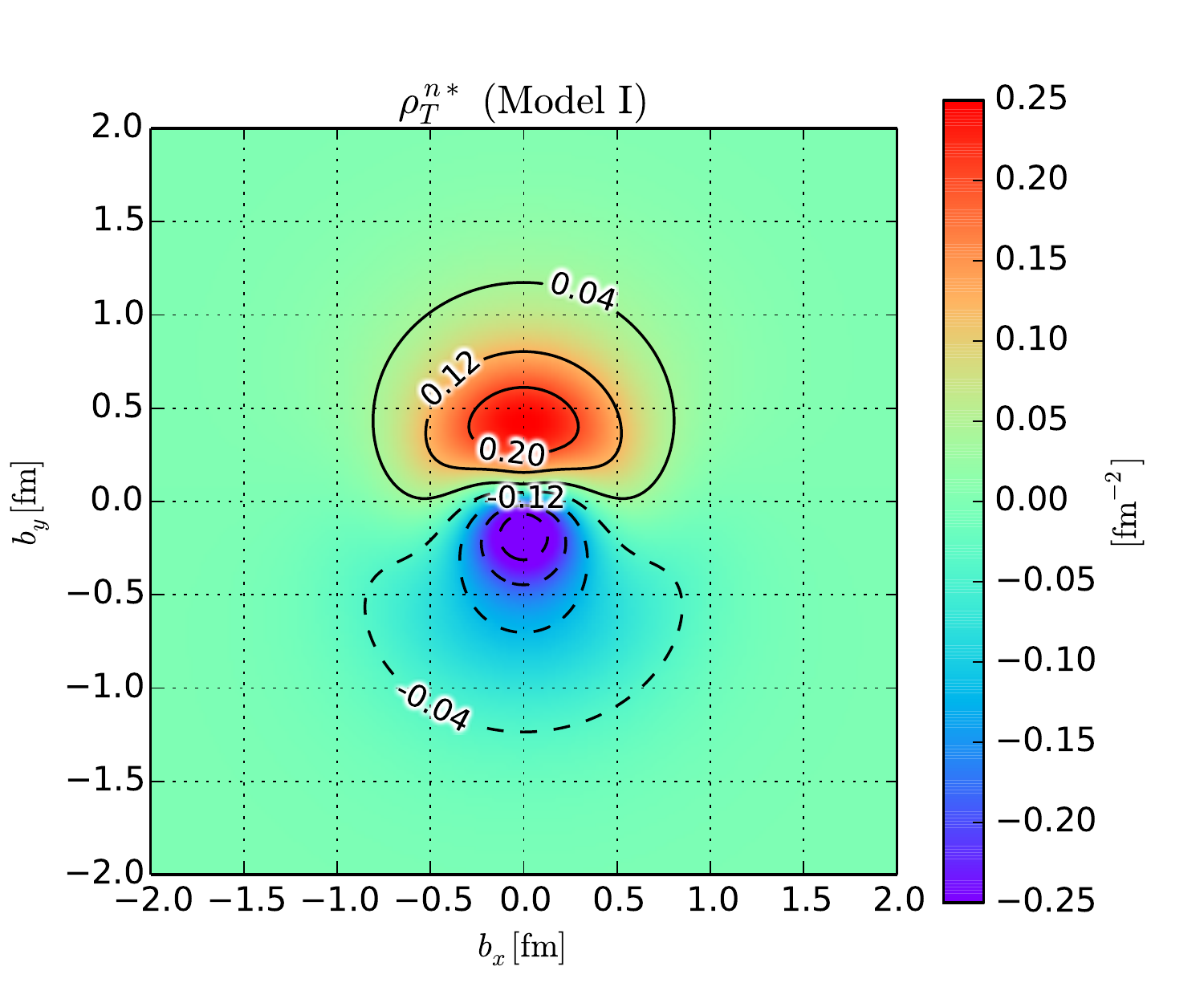}
\includegraphics[scale=0.55]{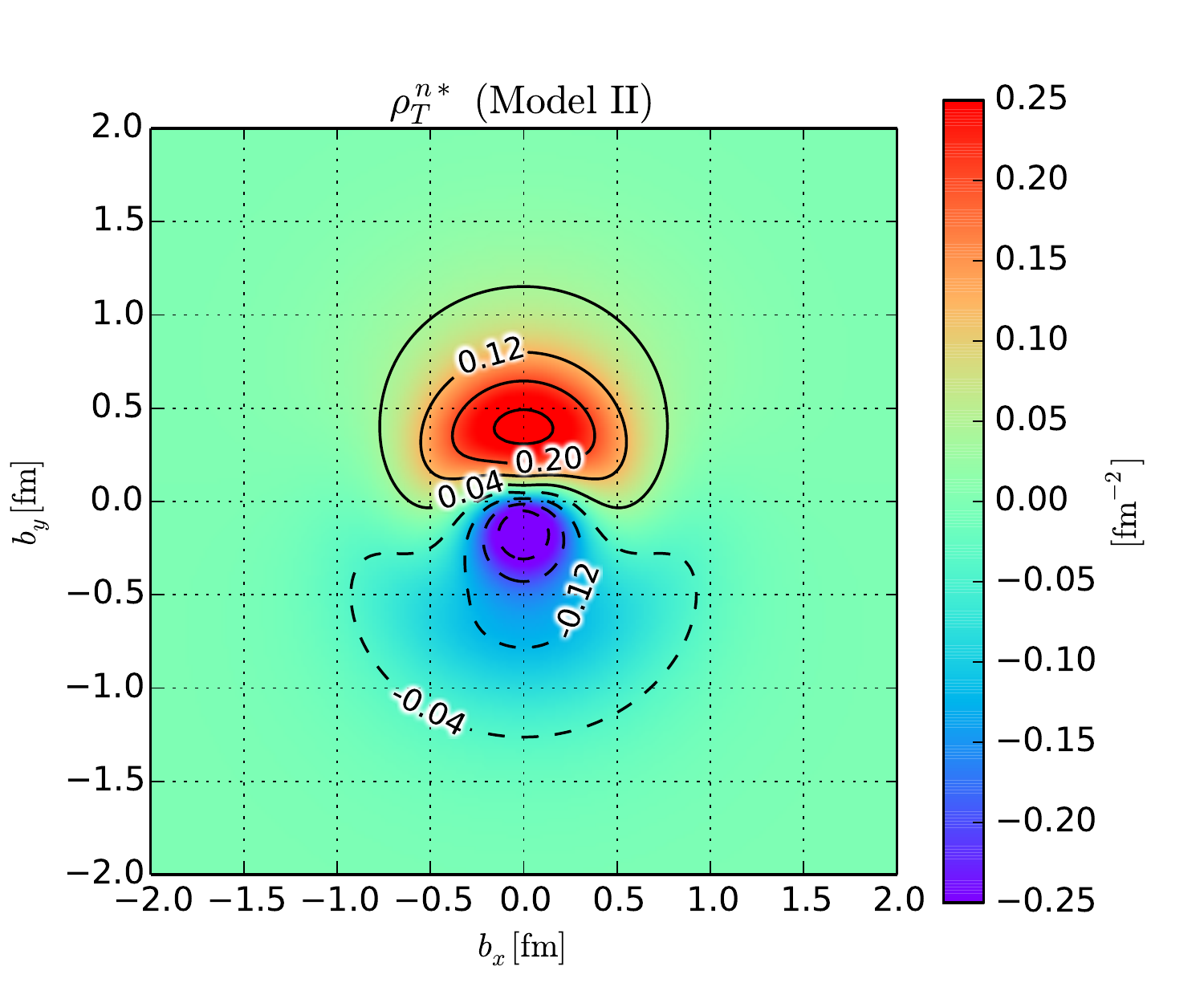}
\par\end{centering}
\caption{Transverse charge densities inside the polarized proton 
(upper panels) and neutron (lower panels) in nuclear medium  
at normal nuclear matter density $\rho_0$ from Model I (left panels) and 
Model II (right panels), respectively.}
\label{fig:polNPTCmn}
\end{figure*}
We are now in a position to discuss the results for the transverse
charge density when the nucleon is polarized. As shown in
Eq.~(\ref{eq:poltrans}), the transverse charge density inside a
polarized nucleon becomes deviated from that inside an unpolarized
nucleon by the second term of the right-hand side of
Eq.~(\ref{eq:poltrans}). Figure~\ref{fig:polNPTCfs} shows the general
feature of the transverse charge densities inside both the polarized
proton (upper-left panel) and the polarized neutron (upper-right
panel). As was already discussed in Ref.~\cite{Carlson:2007xd},
the magnetic field that makes the nucleon polarized along the $x$ axis
produces an induced electric field along the $y$ axis according to
Einstein's theory of special relativity~\cite{Einstein:1905ve}. As a
result, the transverse charge density inside both the polarized proton
is distorted and shifted in the direction of the negative $y$ axis. In
the case of the neutron, the distortion of the corresponding density
is complicated, since the anomalous magnetic moment of the neutron is
negative and the transverse charge density inside an unpolarized
neutron has a different feature, compared with the proton case. Thus,
the negative charged quarks inside a neutron is shifted to the
positive $y$ axis and the positive charged quarks is displaced to the
positive $y$ axis, revealing an asymmetric distortion.  

Figure~\ref{fig:polNPTCmn} illustrates the medium modification of the
transverse charge densities inside both the polarized proton and
neutron. The general behavior of the transverse charge densities
in nuclear medium is very similar to those in free space. However, the
extension of the nucleon size is observed in nuclear medium. Examining
the results shown in Fig.~\ref{fig:polNPTCmn} the effects due to the
polarization of the nucleon are lessened in nuclear medium. This can
be understood from the medium modification of the transverse charge
and magnetization densities inside an unpolarized nucleon as shown in
Fig.~\ref{perpEMchargeDist}. These densities in medium indicate that
the size of the nucleon in medium becomes larger and the effects of
the polarization also get diminished.   

\subsection{EMT form factors and Transverse Energy-Momentum densities}
Let us now discuss the EMTFFs of the nucleon. Since the EMTFFs
correspond to the generalized isoscalar VFFs, 
We do not need to distinguish the proton from the neutron. 
The same results hold for the nucleon embedded into isospin-symmetric
nuclear matter. The situation will change if one introduces the effects of
isospin breaking into the mesonic sector. When one consider more
realistic isospin asymmetric nuclear matter, one has to compute both
the isoscalar and isovector generalized vector FFs. In this case, the
EMTFFs will be regarded only as a part of the GVFFs. In
the present work, we concentrate only on isopin-symmetric nuclear
medium.

The medium-modified EMTFFs of the nucleon have been already
investigated in Ref.~\cite{Jung:2014jja} in detail. Thus, we will
discuss here only the transverse charge and magnetization densities
inside a nucleon, which correspond to the EMTFFs. As shown in
Eq.(\ref{eq:gvffs}), the EMTFFs of the nucleon are identified as the
GVFFs in the NLO, which arise from the second
moments of the vector GPDs. Hence, the transverse charge and
magnetization densities from the EMTFFs of the nucleon can be regarded
as those inside a nucleon to the NLO.  

\begin{figure*}
\includegraphics[scale=0.4]{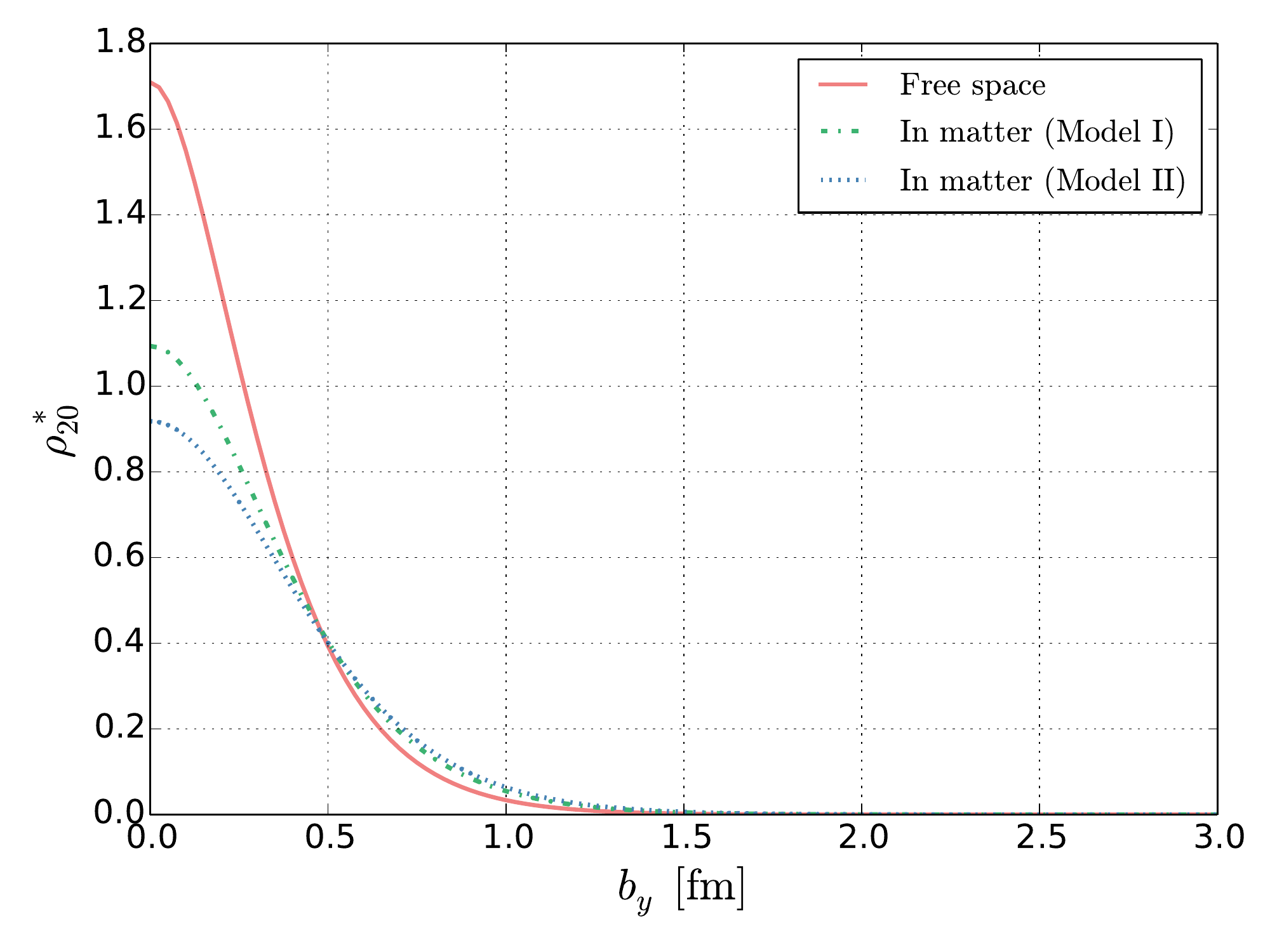}
\includegraphics[scale=0.4]{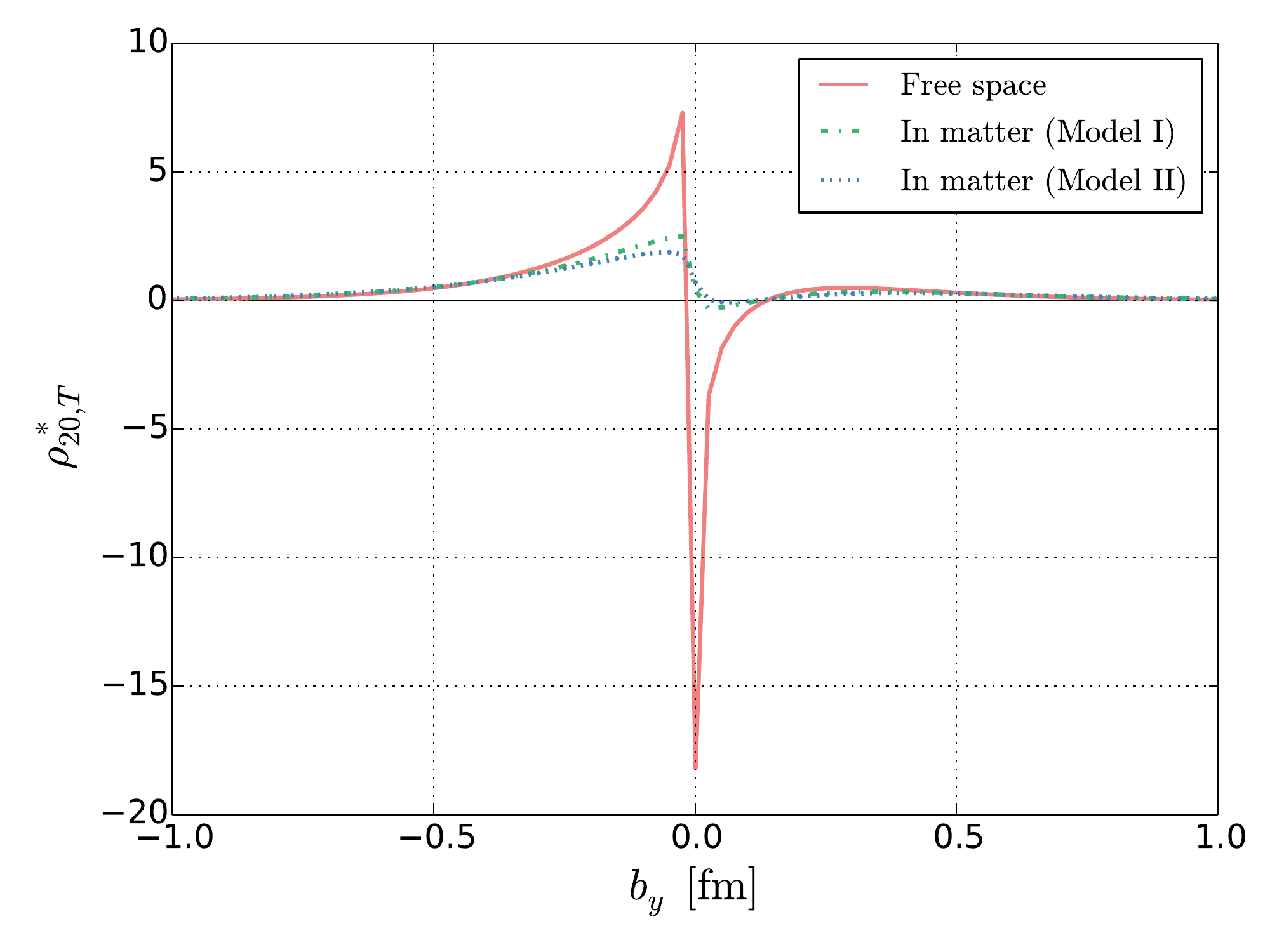}
\caption{The NLO transverse charge densities inside
  the unpolarized nucleon, $\rho_{20}^*$, in the left panel,  and those
  inside the polarized nucleon, $\rho_{20,T}^*$, in the right panel,
  with $b_x = 0$. The solid curve depicts those in free space,
  while the dotted and dotted-dashed ones represent, respectively,
  those from model I and model II in nuclear matter.} 
\label{fig:EMTpunbxFix}
\end{figure*}
Figure~\ref{fig:EMTpunbxFix} draws the NLO transverse charge
densities inside both the unpolarized nucleon (left panel),
$\rho_{20}^*$ and the polarized 
nucleon (right panel), $\rho_{20,T}^*$, with $b_x$ fixed to be
zero. Interestingly, the general feature of $\rho_{20}^*$ is alsmost
the same as $\rho_{\mathrm{ch}}^{p}$ presented in
Fig.~\ref{perpEMchargeDist}. When the nucleon is polarized,
the $\rho_{20,T}$ is changed drastically, as shown in the right panel
of Fig.~\ref{fig:EMTpunbxFix}. However, it can be also 
easily understood as we have discussed previously. The induced
electric field will cause the shift of the positive charged quark to
the negative $y$ direction whereas will translate the negative one to the
positive $y$ axis. The strength of the NLO transverse charge densities
are much decreased in nuclear medium. In particular, the magnitude of
the negative charge is almost suppressed.   

\begin{figure*}
\begin{centering}
\includegraphics[scale=0.55]{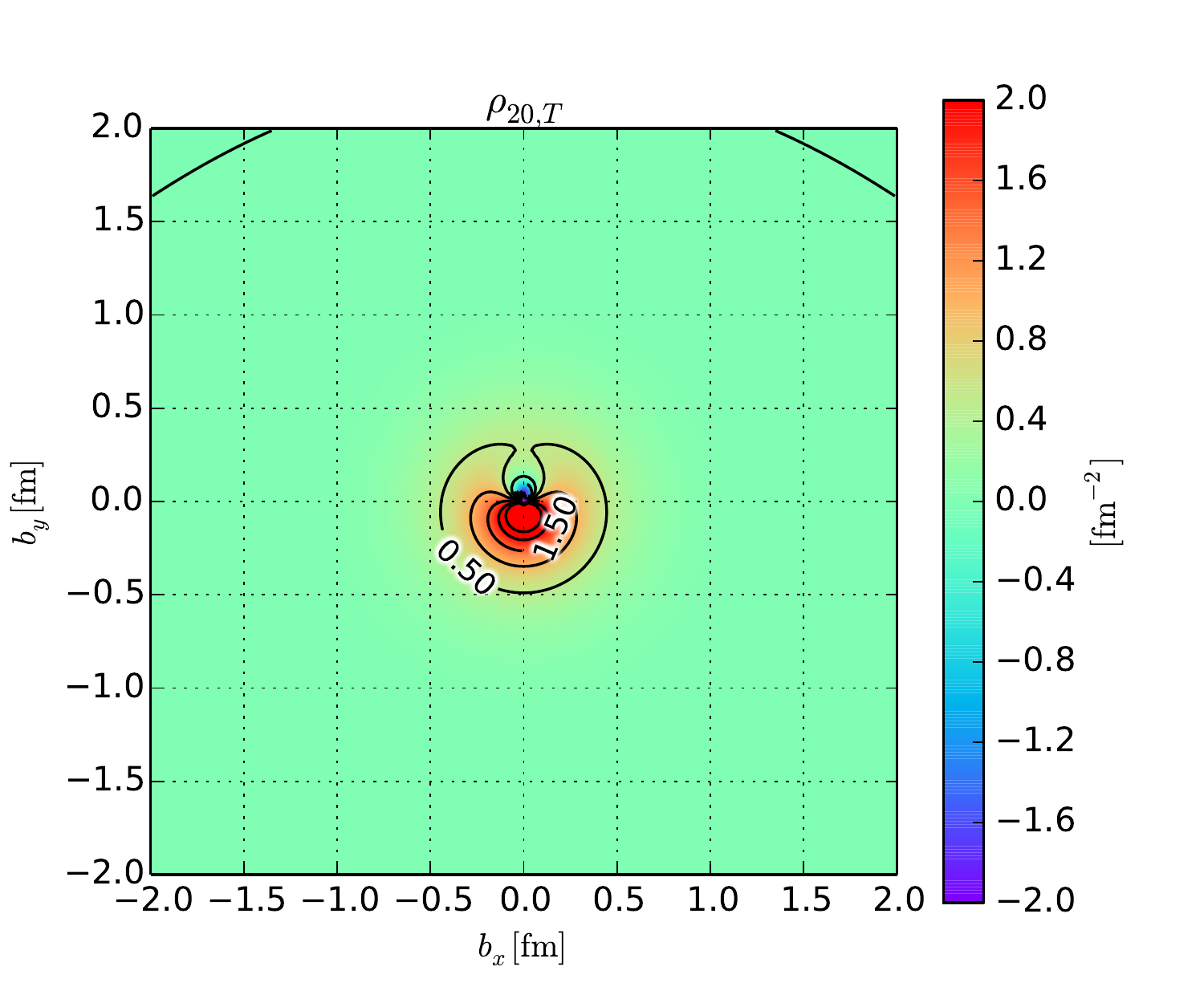}\\
\includegraphics[scale=0.55]{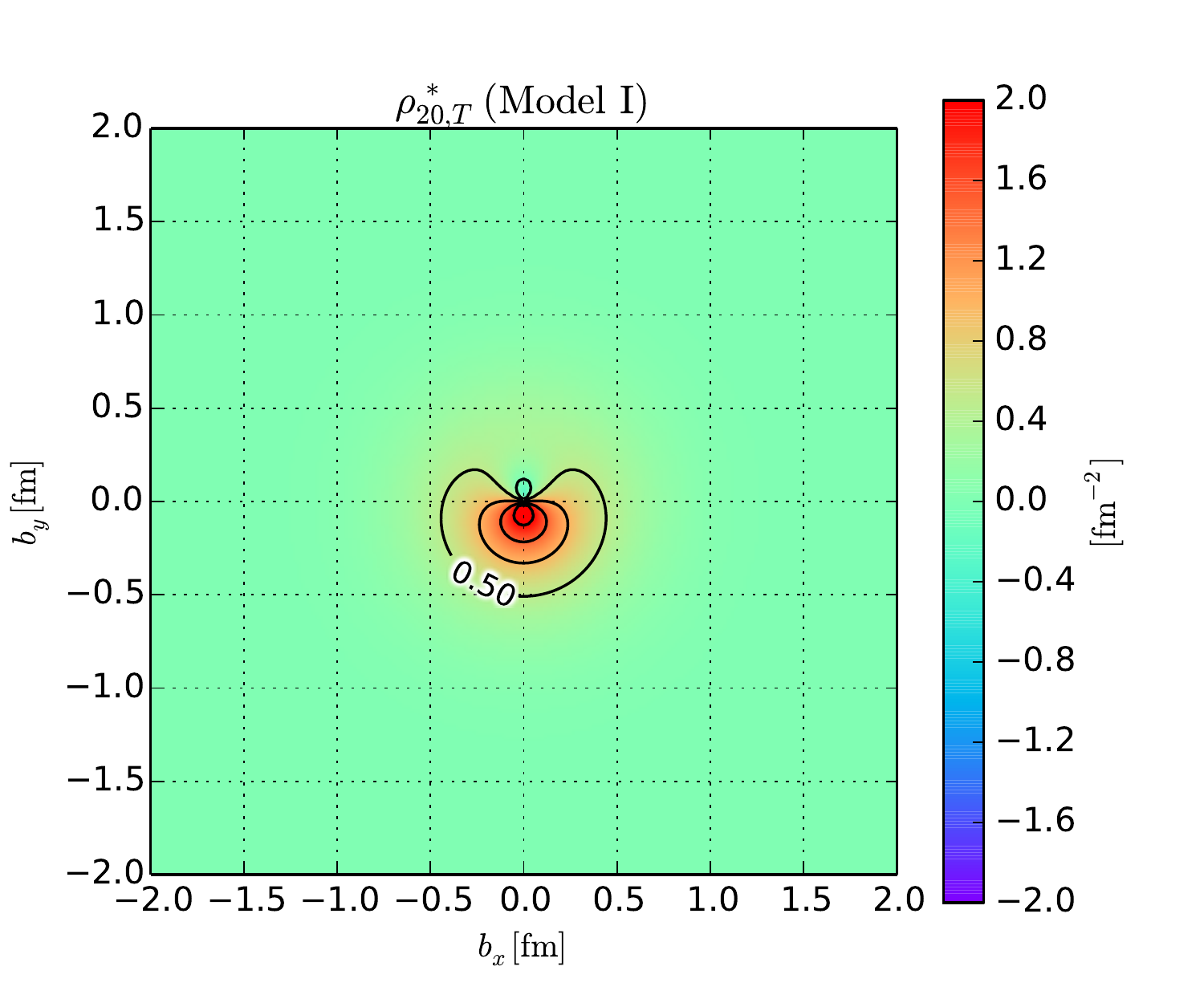}
\includegraphics[scale=0.55]{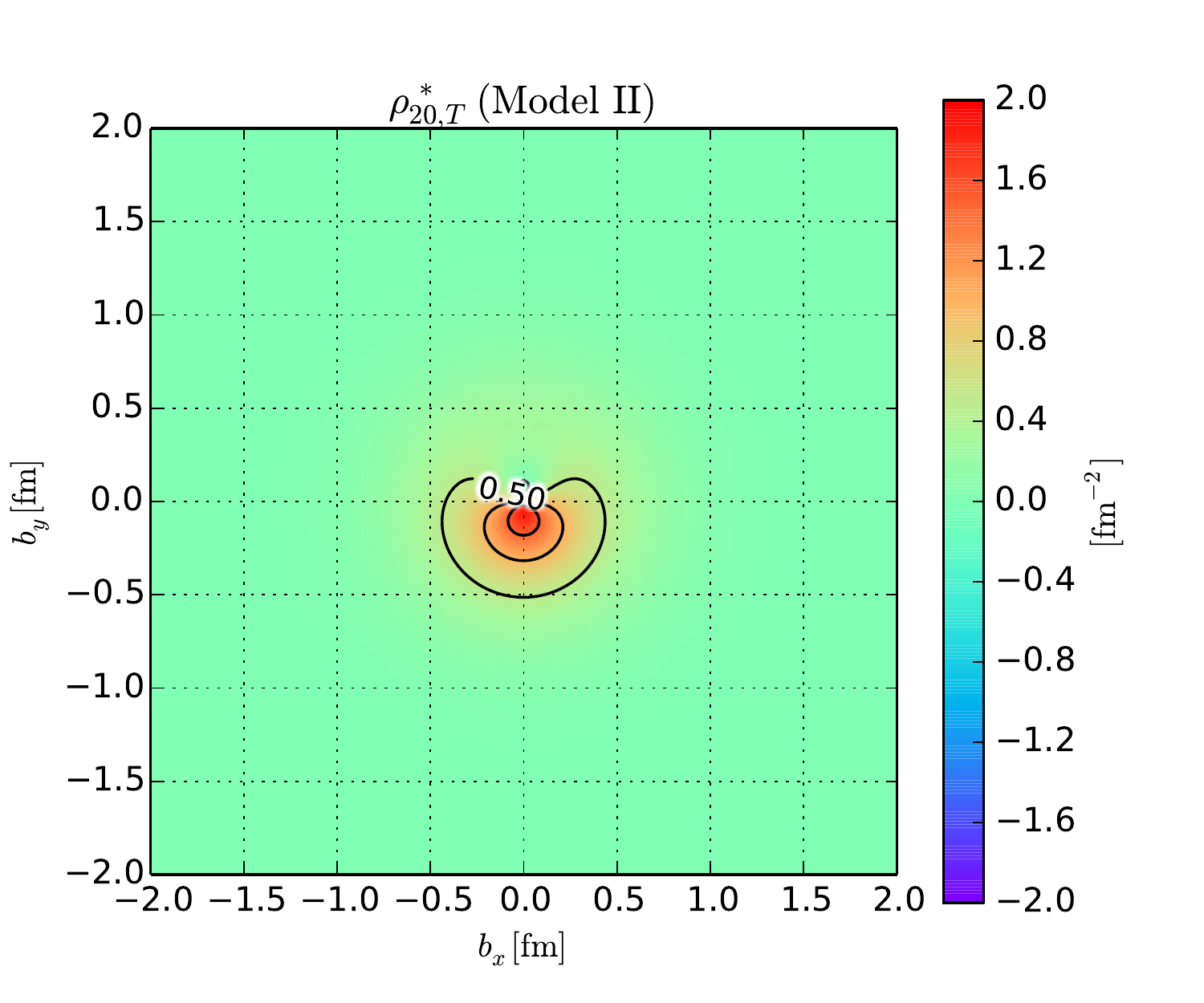}
\par\end{centering}
\caption{The two-dimensional NLO transverse charge
  densities inside the polarized nucleon in free space,
  $\rho_{20,T}$, (upper panel), and those inside the polarized
  nucleon, $\rho_{20,T}^*$, (lower panel). The lower-left panel
  depicts $\rho_{20,T}^*$ from Model I and the lower-right panel draws
those from Model II.}
\label{fig:EMTcdFS}
\end{figure*}

\section{Summary and outlook}
In the present work, we investigated the electromagnetic form factors
of the nucleon in nuclear medium, based on the $\pi$-$\rho$-$\omega$
soliton model. We employed two different models: Model I was
constructed by changing both the $\rho$ meson and the $\omega$ meson
in nuclear matter, while in Model II only the $\rho$ meson undergoes
the change but the $\omega$ meson is intact. This difference yielded
the very different results for the neutron electric form factor. 
We also discussed the transverse charge and magnetization densities
inside both the unpolarized nucleon and the polarized nucleon. The
densities showed that the nucleon swells in nuclear matter, which was
also the case in the medium-modified Skyrme model. The effects of the
nucleon polarization turned out to be lessened in nuclear matter.
Finally, we presented the results for the next-to-leading order
transverse charge densities obtained from the energy-momentum tensor
form factors or the generalized vector form factors of the nucleon.

Based on the $\pi$-$\rho$-$\omega$ soliton model, it is also of great
interest to study the spin problem of the nucleon, in particular, the
spin densities of the nucleon~\cite{Diehl:2005jf}. While the model
does not contain any quark degrees of freedom, it is still possible to
study the quark spin distributions inside a nucleon. The present work
will shed light on the spin structure of the nucleon from a
complementary view point and furthermore on the changes of its spin
structure in nuclear medium. The corresponding work is under way.  

\section*{Acknowledgments}
This work is supported by the Basic Science Research
Program through the National Research Foundation (NRF) of Korea funded 
by the Korean government (Ministry of Education, Science and
Technology), Grant Numbers~2011-0023478 (JHJ and UY) and 
NRF-2013S1A2A2035612 (HChK). JHJ also acknowledges a partial
support by the "Fonds zur F\"orderung der wissenschaftlichen Forschung
in \"Osterreich via FWF DK W1203-N16."

\end{document}